\documentclass[fleqn,10pt]{wlscirep}
\usepackage[utf8]{inputenc}
\usepackage[T1]{fontenc}
\usepackage{bm}

\usepackage{changes}

\title{Physics-Driven Learning for Inverse Problems in Quantum Chromodynamics}

\author[1]{Gert Aarts}
\author[2]{Kenji Fukushima}
\author[3]{Tetsuo Hatsuda}
\author[4]{Andreas Ipp}
\author[5]{Shuzhe Shi}
\author[3,*]{Lingxiao Wang}
\author[6,7]{Kai Zhou}
\affil[1]{Department of Physics, Swansea University, SA2 8PP, Swansea, United Kingdom}
\affil[2]{Department of Physics, The University of Tokyo, 7-3-1 Hongo, Bunkyo-ku, Tokyo, 113-0033, Japan}
\affil[3]{Interdisciplinary Theoretical and Mathematical Sciences Program (iTHEMS), RIKEN, Wako, Saitama 351-0198, Japan}
\affil[4]{Institute for Theoretical Physics, TU Wien, Wiedner Hauptstraße 8-10/136, A-1040 Vienna, Austria}
\affil[5]{Department of Physics, Tsinghua University, Beijing 100084, China}
\affil[6]{School of Science and Engineering, The Chinese University of Hong Kong, Shenzhen (CUHK-Shenzhen), Guangdong, 518172, China}
\affil[7]{Frankfurt Institute for Advanced Studies, Ruth Moufang Strasse 1, D-60438, Frankfurt am Main, Germany}
\affil[*]{lingxiao.wang@riken.jp}

\begin{abstract}
The integration of deep learning techniques and physics-driven designs is reforming the way we address inverse problems, in which accurate physical properties are extracted from complex data sets. This is particularly relevant for quantum chromodynamics (QCD), the theory of strong interactions, with its inherent limitations in observational data and demanding computational approaches. This perspective highlights advances and potential of physics-driven learning methods, focusing on predictions of physical quantities towards QCD physics, and drawing connections to machine learning(ML). It is shown that the fusion of ML and physics can lead to more efficient and reliable problem-solving strategies. Key ideas of ML, methodology of embedding physics priors, and generative models as inverse modelling of physical probability distributions are introduced. Specific applications cover first-principle lattice calculations, and QCD physics of hadrons, neutron stars, and heavy-ion collisions. These examples provide a structured and concise overview of how incorporating prior knowledge such as symmetry, continuity and equations into deep learning designs can address diverse inverse problems across different physical sciences.
\end{abstract}
\begin{document}

\flushbottom
\maketitle

\thispagestyle{empty}


\section*{Introduction}

Quantum chromodynamics (QCD) is the fundamental theory describing the strong interactions of quarks and gluons, building hadrons such as protons and neutrons~\cite{Gross:2022hyw}. Characterised by its non-perturbative nature at low energies, QCD physics derives phenomena such as confinement (permanent binding of quarks and gluons within hadrons) and asymptotic freedom (weaker quark interactions at high energies). QCD physics is crucial for understanding nuclear and extreme matter, which is promising to be achieved through first-principle lattice calculations~\cite{Gattringer:2010zz}, compact star observations~\cite{Baym:2017whm}, and relativistic nuclear collision experiments~\cite{Yagi:2005yb}.

The quest to explore QCD physics and decode its associated phenomena involves many challenging inverse problems, which need to determine causes or parameters from consequent observations~\cite{tanaka2021deep,Zhou:2023pti}. Unlike forward problems, which predict outcomes from known factors, inverse problems start from results to do reverse engineering~\cite{kaipio2006statistical}. They are essential in fields such as medical imaging, geophysics and astrophysics, where direct measurements are impractical. Inverse problems in QCD physics involve identifying strong interaction properties from complex measurements. Examples include extracting hadron spectral functions~\cite{Asakawa:2000tr} from lattice observables, reconstructing dense matter equations of state (EoSs) from compact star observations~\cite{Yunes:2022ldq}, and identifying quark-gluon plasma (QGP) properties from heavy-ion collision (HIC) experiments~\cite{Zhou:2023pti}.

Machine learning (ML) techniques, as a modern branch of artificial intelligence (AI), are becoming increasingly important in QCD physics~\cite{Boehnlein:2021eym,Zhou:2023pti}, providing tools to unravel intricate patterns and extract structures from complex data sets~\cite{LeCun:2015pmr,bishop2023deep}. Bayesian inference can deduce causal parameters from uncertain observations, and deep neural networks (DNNs) are trained to learn physical properties from well-prepared data. Recently, advanced developments in deep learning are pushing the boundary to a high-accuracy frontier. Physics-driven learning methods integrate physical rules and constraints into deep learning frameworks~\cite{Raissi:2017zsi,2021arXiv210905237T,Zhou:2023pti}, improving deep models to produce physically meaningful and accurate results. These burgeoning methods address inverse problems towards QCD by incorporating physics priors directly into neural network architectures or learning processes. This ensures solutions adhere to the physics world, reducing the parameter space and improving the accuracy, which is still missing in current large models.

Physical concepts have been fundamental in the development of ML~\cite{Carleo:2019ptp}. Mathematical frameworks of physics and problem-solving techniques have inspired algorithms and models of deep learning, including e.g., energy-based models~\cite{lecun2006tutorial}, maximum entropy thermodynamics~\cite{Jaynes:1957zza}, non-equilibrium stochastic dynamics \cite{DBLP:journals/corr/Sohl-DicksteinW15},
and physics-inspired neural network design and optimisation~\cite{muller1995neural}. The rapidly revolutionary physics-driven learning paradigm is both helpful in solving inverse problems, and will in turn feed AI innovations, such as developing more physically meaningful generative models and aligning deep models to the physical world.

\section*{Methodology}

\subsection*{Statistical Inference and Deep Learning}
In solving inverse problems towards QCD physics, statistical inference and deep learning (DL) emerge as indispensable tools, each contributing uniquely to the resolution of the involved challenges.

Statistical inference provides a rigorous probabilistic framework for understanding and quantifying uncertainties inherent in QCD physics. Powered with Bayesian inference (BI), one systematically incorporates prior knowledge to update beliefs on the cared physics based upon observations, and the posterior distribution $ p(\theta | D) $ of the parameters $\theta$ given the observations $D$ is obtained using Bayes' theorem~\cite{murphy2012machine}, $p(\theta | D) = p(D | \theta) p(\theta)/p(D)$, where $ p(D | \theta) $ is the likelihood, $ p(\theta) $ is the prior, and $ p(D) $ is the evidence. This allows for the integration of prior theoretical models or simulated data, providing a coherent method to update the physics causal parameters as new data become available, enabling also the estimation of parameter uncertainties and the construction of credible intervals.

Deep learning ~\cite{LeCun:2015pmr}, with its capacity to model complex, non-linear relationships using DNNs, offers a complementary approach for inverse problems. The network learns to infer the latent parameters $\theta$ from observations $D$, performing inverse engineering, {which usually works under the principle of maximum likelihood estimation. With this objective, one can choose different architectures, such as convolutional neural networks (CNNs)~\cite{7949028}, recurrent neural networks~\cite{AMIN2007491}, residual networks~\cite{behrmann2019invertible}, graph neural networks~\cite{qi2017pointnet} and transformers~\cite{Tomiya:2023jdy} for various inverse problems. In the context of QCD, DL can be employed to approximate the inverse mapping from observations to the underlying physics. This is achieved by training DNNs on datasets generated from theoretical models or experimental simulations. The network learns to infer the latent parameters $\theta$ from observations $D$, effectively performing inverse engineering.

The integration of DL and BI reinforces the strengths of both, which applies DNNs within Bayesian perspective, thus combines the flexibility of DNNs with the probabilistic rigour of BI. This incorporates uncertainty into the network weights, allowing for the estimation of uncertainty of the predictions.

\begin{figure}[htbp!]
\centering
\includegraphics[width=0.9\linewidth]{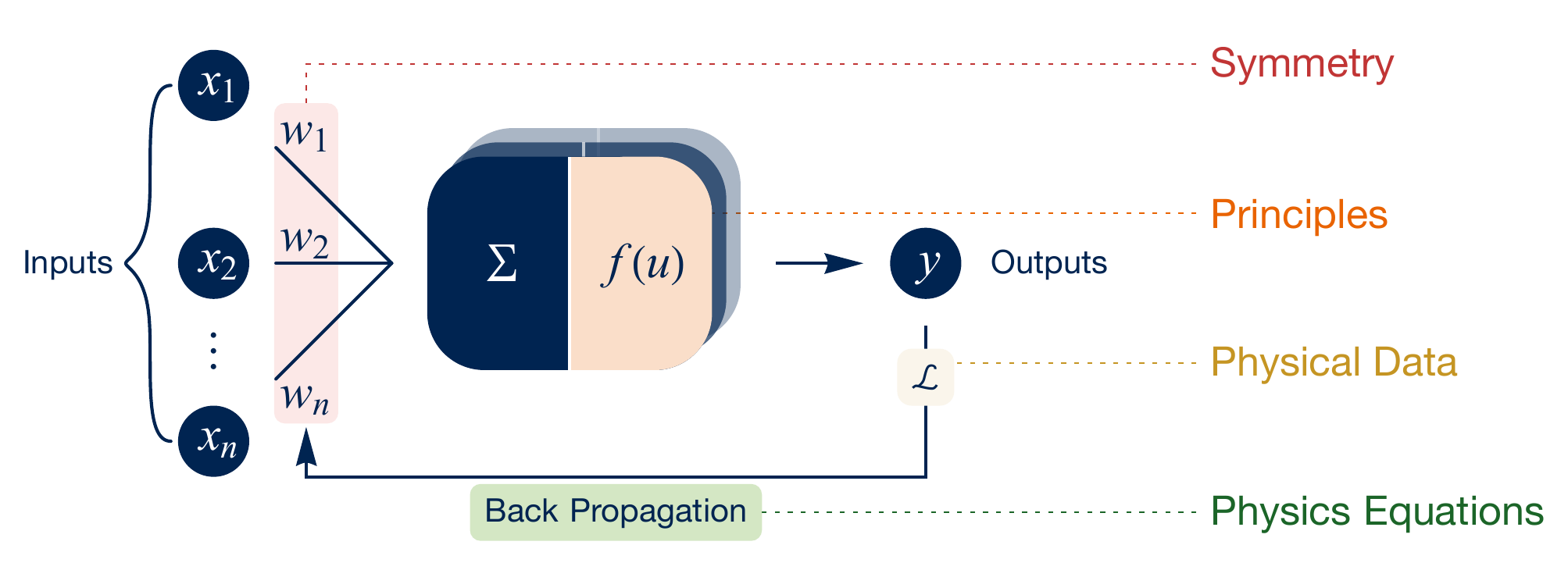}
\caption{Physics-driven deep learning. In a deep neural network model, the weights, $\{w\} = w_1, w_2, \dots, w_n$, connect the inputs, $\{x\} = x_1, x_2, \dots, x_n$, and the outputs, $y$, with summation $\Sigma$ and non-linear activation functions $f(u)$. In a single layer, the equation can be simplified as $y = f(\Sigma_{i=1}^n x_i w_i)$. The \textit{symmetries} can be encoded within the weights, and other \textit{principles} can be represented by different activation functions. Due to its differentiable properties, \textit{physics equations} can be explicitly utilised in the back propagation (BP) algorithm. The \textit{physical data} provides guidance for the outputs from deep models when computing the loss functions.}
\label{fig:phy-dl}
\end{figure}

\subsection*{Physics-Driven Learning}

Solving inverse problems involves determining the causes from observations, which can be particularly difficult when data is incomplete or noisy. These problems are often ill-posed, meaning that solutions may not exist, not unique, or not stable~\cite{kaipio2006statistical}. Advanced deep learning techniques and regularisation methods are used to obtain meaningful solutions. Integrating physics priors, insights, and properties into machine learning methods presents a promising strategy for tackling the problems more effectively~\cite{raissi2019physics,Zhou:2023pti}. One newly-developing paradigm aims to introduce physical knowledge explicitly to solve inverse problems, which is named as \textit{physics-driven learning}. The physical knowledge encompasses symmetries, principles, well-developed physics equations and simulations, as Figure~\ref{fig:phy-dl} shows. These elements will also facilitate the development of more advanced and reliable deep learning approaches.

\textbf{Symmetry}, a cornerstone in modern physics, can be employed to improve learning performance explicitly. In principle, embedding symmetries introduces a scheme for sharing parameters in deep models, thus reducing the number of parameters and preventing over-fitting~\cite{2019arXiv190408991M}. Symmetries such as \textit{translational, rotational}, and \textit{permutation invariance} can be inherently incorporated into network architectures, e.g., CNNs~\cite{zhang1988shift,bishop2023deep}, Euclidean Neural Networks (e3nn)~\cite{2022arXiv220709453G} and Graph Neural Networks (GNNs)~\cite{Shlomi_2021}. In QCD physics, gauge invariance and equivariance are particularly important for embedding in deep models~\cite{Kanwar:2020xzo,Favoni:2020reg,Cranmer:2023xbe}.

Moreover, other \textbf{principles} such as \textit{causality, continuity}, \textit{positive definiteness} and \textit{asymptotic behaviours}, etc., play crucial roles in ensuring that the solutions are physically meaningful. These properties can be also embedded into architecture design, either through customised loss functions or specialised activation functions~\cite{Han:2021kjx,Shi:2022yqw}, ensuring that the output respects these fundamental constraints.

\textbf{Physics equations}, concertised by differential equations governing systems, can also serve as priors. These laws provide essential constraints that guide the learning process, ensuring that solutions adhere to physical realities. Different from the physics-informed neural networks(PINNs)~\cite{karniadakis2021physics}, here physics equations can be encoded into the optimisation explicitly. For instance, ordinary/partial differential equations (ODEs/PDEs) are automatic differentiable~\cite{baydin2018automatic}, which can be encoded into the forward process and their gradients can be computed in reverse optimisations for training, leading to more robust and physically controllable outcomes~\cite{2021arXiv210905237T}.

\textbf{Physical data}, whether obtained from experiments or simulations~\cite{Cranmer:2019eaq}, serves as a form of regularisation, which assists in aligning the model outputs with physical truths~\cite{wang2023scientific}. In particular, for ill-posed inverse problems and those that necessitate initial verification of the existence of the inverse mapping, this regularisation can not only ground the learning process in reality but also mitigate the risk of obtaining non-physical solutions.

Designing deep learning models with specific physics knowledge can further enhance their capability and effectiveness~\cite{Carleo:2019ptp,jalali2022physics}. For example, incorporating domain-specific knowledge into the design of deep generative models. In which, \textbf{inverse modelling} underlying probability distributions are the backbone of generative models~\cite{bishop2023deep}. Whether starting from exact likelihood estimation or not, one can approximate the underlying physical distributions and generate reliable samples. These models can benefit from existing and well-verified physics knowledge.

\section*{QCD Physics}

\subsection*{Lattice QCD}

Lattice QCD (LQCD) provides a first-principle, non-perturbative approach to study the strong interactions.\cite{Gattringer:2010zz} Progress over the past 50 years has been driven by a combined improvement of algorithms and increase in computational power, executing simulations on the largest supercomputers. The workhorse of LQCD is Markov Chain Monte Carlo, to generate large ensembles of field configurations in four-dimensional spacetime. The low-temperature, moderate-to-high-density region of the phase diagram is still out of bounds due to the sign problem: at nonzero baryon chemical potential the quark determinant is complex and importance sampling cannot be used.\cite{Aarts:2015tyj}

ML provides a new tool to investigate and enhance LQCD simulations and indeed ML is already used to study many aspects, including configuration generation, observable measurement and analysis.\cite{Boyda:2022nmh} Many applications are still exploratory and most are developed in theories easier than QCD; nevertheless there is promise and we will discuss selected examples here. An important aspect is precision and exactness: to compete with or improve upon well-established methods, it is necessary to demonstrate that ML-driven algorithms can deliver high-precision results with controllable systematic and statistical uncertainties.

\begin{figure}[htbp!]
\centering
\includegraphics[width=0.9\linewidth]{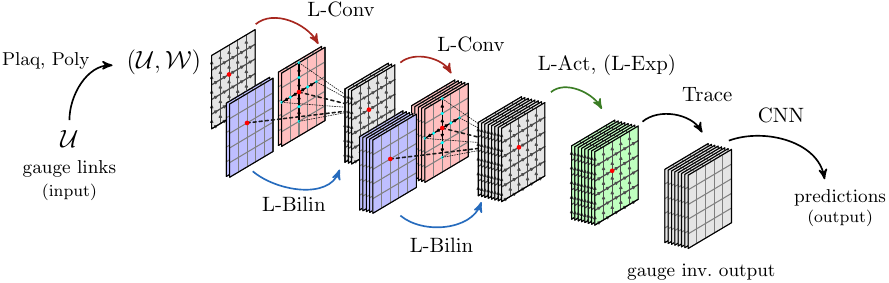}
\caption{Example of an L-CNN. Each layer preserves gauge equivariance on the lattice, such that convolutional, bilinear, activation or exponentiation layers can be combined for the use in inverse problems where the preservation of this symmetry is crucial. Contrary to conventional CNNs, they are robust to random and adversarial gauge transformations. Image from \cite{Favoni:2020reg}.}
\label{fig:l-cnn}
\end{figure}

Before starting an LQCD simulation, parameters in the lattice action need to be determined.
To minimise discretisation effects, one can employ {\em perfect} or {\em fixed point actions}, which are classically free of discretisation errors but come with an infinite number of couplings in principle.\cite{Hasenfratz:1993sp,DeGrand:1995ji}
In pioneering works, ML has been used to learn action parameters~\cite{Shanahan:2018vcv} with the aim of improving lattice simulations \cite{Nagai:2020jar} and gaining explainable insight into lattice systems~\cite{Blucher:2020mjt}. Generically, gauge symmetry had to be learned by such networks, or was included through manually selected small Wilson loop structures. A more scalable approach to learning fixed point actions has been enabled recently \cite{Holland:2023ews,Holland:2024muu} by using lattice-gauge-equivariant CNNs (L-CNNs) to determine parametrisations which are shown to be superior to the ones considered previously. L-CNNs are essential in achieving this, in that they provide\cite{Favoni:2020reg} a very general formulation of a lattice action, e.g.\ as functions of arbitrarily sized Wilson loops, while maintaining exact gauge symmetry in the network architecture. 

Most attention has gone to ensemble generation, due to its link with generative AI. Early applications \cite{Zhou:2018ill,Pawlowski:2018qxs} employed {\em Generative Adversarial Networks} to generate field configurations in a two-dimensional scalar field theory. {\em Normalising flow} is by now the best developed approach and been reviewed extensively elsewhere.\cite{Cranmer:2023xbe,Kanwar:2024ujc}
{\em Diffusion models} (DMs) are of interest for a number of reasons: the underlying idea comes from physics\cite{DBLP:journals/corr/Sohl-DicksteinW15,DBLP:journals/corr/abs-2011-13456} and the stochastic updates are easily understood by lattice field theorists. Indeed the relation between DMs and stochastic quantisation, \cite{Parisi:1980ys} a well-known technique to generate quantum field configurations via a stochastic process in a fictitious time dimension, has been pointed out recently~\cite{Wang:2023exq}, and first applications to scalar and U(1) gauge field theories in two dimensions have been implemented~\cite{Wang:2023exq,Wang:2023sry,Zhu:2024kiu}. In contrast to normalising flow, DMs learn from configurations previously generated via any alternative approach, such as hybrid Monte Carlo (HMC), but the trained DM can subsequently be incorporated in the Markov Chain to increase the size of ensembles. 
Further connections of DMs to the path integral \cite{Hirono:2024zyg} and renormalisation group flows \cite{Cotler:2023lem} have been pointed out as well. 

A quite different generative approach employs the {\em inverse renormalisation group} (IRG), originally proposed for spin systems.\cite{PhysRevLett.89.275701} Here the idea is to learn a transformation that undoes a standard RG transformation, using transposed convolutions.\cite{Bachtis:2021eww} If the inverse transformation is local, it can be applied over and over again, generating larger lattices, closer to criticality. This idea has first been applied to scalar fields in two dimensions\cite{Bachtis:2021eww} and also to a hard-to-simulate disordered system, the three-dimensional Edwards-Anderson model.\cite{Bachtis:2023ykk} The crucial step to make this work is to design a ``good'' RG step, which relies on the understanding of the physics in the critical regime.

Finally, an area in which ML and lattice field theory (LFT) intersect in a different fashion relies on the observation that both LFT and ML models describe systems with many fluctuating degrees of freedom. While the starting points are clearly different --- QCD is a fundamental theory, with a probability distribution (or action) fixed by symmetries and renormalisation conditions; ML models learn or approximate distributions mostly driven by data --- there are similarities whose exploration can have mutual benefits. A prominent example is the presence of local (gauge) symmetries, leading to gauge-equivariant networks for lattice systems, \cite{Favoni:2020reg,Lehner:2023bba,Aronsson:2023rli,Lehner:2023prf} extending concepts introduced earlier in the ML literature.\cite{DBLP:journals/corr/CohenW16,DBLP:journals/corr/abs-1902-04615}.  
ML can be used to learn order parameters and determine the phase structure of the simulated theory,\cite{Carrasquilla_2017} which has been extended to lattice gauge theories.\cite{Wetzel:2017ooo,Boyda:2020nfh}
It may also be useful to analyse ML models as (quantum) field theories, building on the relation between deep NNs in the limit of infinite width to Gaussian processes and free fields.\cite{lee2018deep,Halverson:2020trp} Alternatively, one can add LFT interactions on the nodes\cite{Bachtis:2021xoh} or use an LFT analysis to provide a new perspective, as demonstrated in the case of Gaussian Restricted Boltzmann Machines~\cite{Aarts:2023uwt} or via the relation to Dyson Brownian motion and random matrix theory~\cite{Aarts:2024wxi}.
The merger of physics intuition with the framework of ML is expected to deepen understanding going forward.

\subsection*{Hadron Physics}
Much understanding of the subatomic world is rooted in hadron physics. Recently, machine learning techniques have become useful tools to further this understanding as they help to understand hadron spectra, search for exotic hadrons and study hadron interactions.

The theoretical extraction of hadron spectra from LQCD, which is an inverse problem, is a challenging task because there is only a finite amount of LQCD data with statistical noise. Apart from the standard method of extracting hadron masses using long-range temporal correlations, the Maximum Entropy Method~\cite{Asakawa:2000tr,Rothkopf:2019ipj}, based on Bayesian inference, provides a robust framework for the extraction of hadron spectral functions from LQCD data using information entropy for regularization. Recently, automatic differentiation methods (Figure~\ref{fig:ad}) based on maximum likelihood estimation have been developed. The spectral function is first represented by a neural network Ansatz, then the difference between the predictions and the real observations is computed as a loss function $\mathcal{L}$, whose minimization can be back-propagated to optimize parameters $\{\theta\}$ in neural networks, as $\partial \mathcal{L}/\partial \theta$. In the forward process, the physical integral is explicitly encoded as a sum to compute the predictions. The principles that the spectral function has to be continuous and positive are also explicitly incorporated into the flexible neural network representations~\cite{Wang:2021jou,Shi:2022yqw}.
\begin{figure}[htbp!]
\centering
\includegraphics[width=0.7\linewidth]{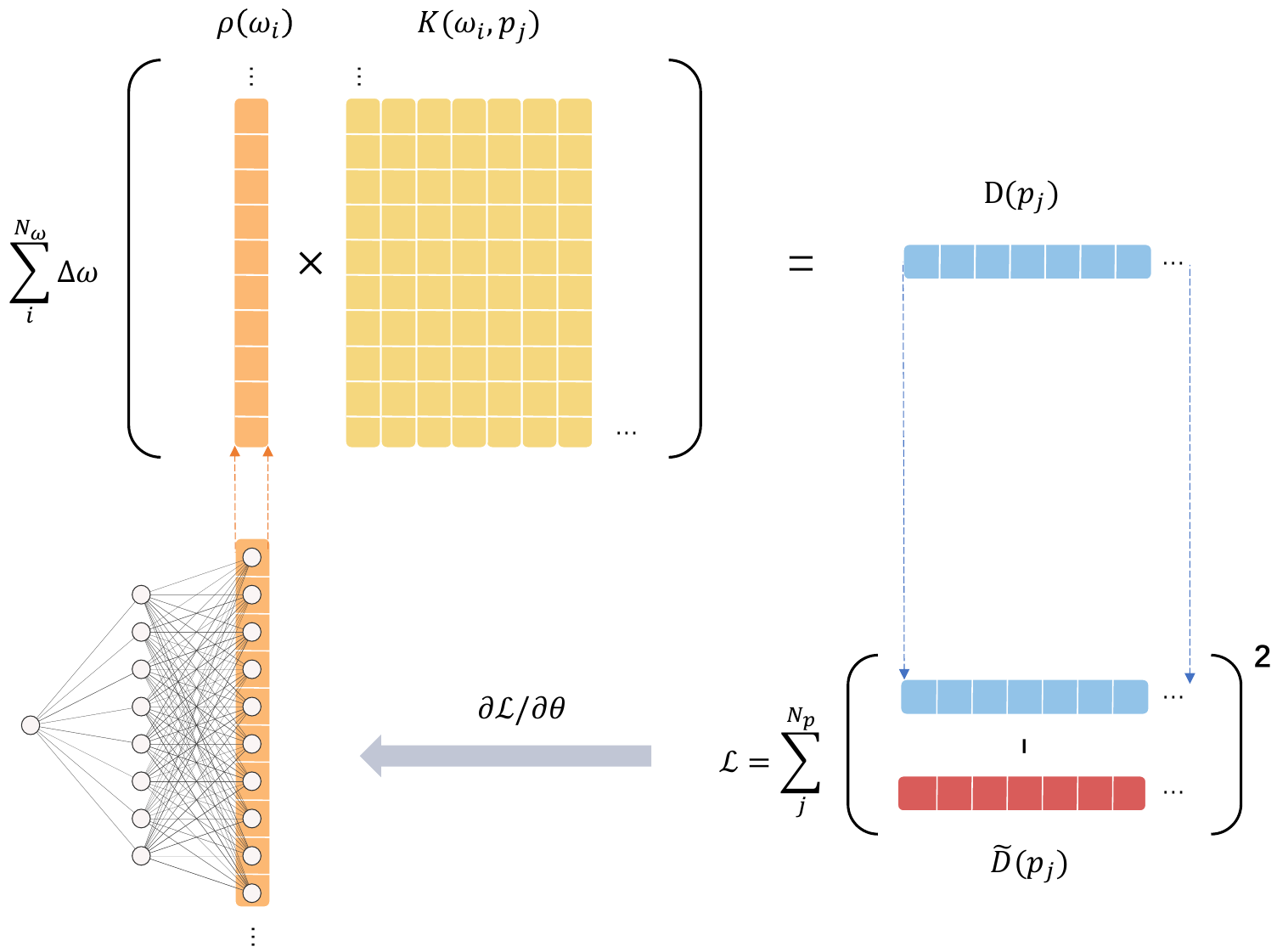}
\caption{Automatic differential framework to reconstruct spectral functions from observations. Neural networks have outputs as a list representation of spectrum $\rho(\omega_i)$. The convolutional operation between $\rho(\omega_i)$ and kernel function $ K(\omega, p)$, gives the predicted observations $D(p)$, as $D(p_j) = \sum_i^{N_\omega} \Delta \omega \rho(\omega_i) K(\omega_i, p_j)$. The difference between real observations and predicted one is utilised to compute the loss function, $\mathcal{L} = \sum_j^{N_p}(D(p_j) - \tilde{D}(p_j))^2$, for optimising the weights $\{\theta\}$ of neural networks, with gradient $\partial\mathcal{L}/\partial \theta$. The activation functions of the neural network can be set as \textit{Softplus} to meet the positive definition principle.}
\label{fig:ad}
\end{figure}

The exploration of exotic hadrons has become an exciting frontier in hadron physics~\cite{Guo:2017jvc}. Exotic states, including tetraquarks, pentaquarks and other multiquark states challenge the traditional meson and baryon picture. Candidates of exotic hadrons are often found experimentally as peaks near certain decay thresholds, and it is important to identify theoretically whether they correspond to bound, virtual or resonant poles in the complex plane. Bound poles represent stable systems in which particles are permanently bound, virtual poles reflect transient interactions that affect scattering without forming stable states, and resonant poles represent unstable, short-lived states that decay into other particles. One can detect such poles from partial wave amplitudes using deep neural networks~\cite{Sombillo:2020ccg,Sombillo:2021rxv} by taking into account the general properties of the $S$-matrix, such as analyticity and unitarity. ML studies on the properties of the narrow pentaquark $P_c(4312)$ ~\cite{JPAC:2021rxu} as well as other exotic states~\cite{Ng:2021ibr} are currently an active field or research, and neural network wave-functions for hadrons~\cite{Keeble:2019bkv,Adams:2020aax} may help to identify and characterize these exotic states.

Hadronic interactions play a crucial role not only in the investigation of the structure of exotic hadrons but also in the understanding of how atomic nuclei form and of the physics inside neutron stars~\cite{Baym:2017whm}. One approach to study hadron interactions is the HAL QCD method~\cite{Ishii:2006ec,Aoki:2012tk,Aoki:2023qih}, which has been proposed  as a way to build effective potentials between hadrons from their spatial correlations (equal-time Nambu-Bethe-Salpeter amplitude) measured on the lattice. This approach could bridge the gap between LQCD theory and experimental data~\cite{ALICE:2020mfd,Lyu:2024kwz}. Deriving the potential from spatial correlations in LQCD is also an inverse problem with which physics-driven learning can help: symmetric neural networks, trained on LQCD data with constraints of asymptotic behaviours, can model general non-local potentials in the Schr\"odinger equation~\cite{Shi:2021qri,Wang:2024ykk}.

\subsection*{Neutron Stars}

We would commonly encounter ill-posed problems in physics systems that can be probed only by rare events such as astronomical phenomena. One great and unsolved puzzle in physics is how extremely compressed matter should look like.  Since the total number of baryons is conserved in the Standard Model (except for rare tunnelling processes), the highly compressed matter has a large density.  In ordinary terrestrial environments the density of compressed matter is saturated by that at the core of heavy nuclei, i.e., the nuclear saturation density $\simeq 2.7\times 10^{14}\;\mathrm{g/cm}^3$. To compress matter further, we should consider extreme environments as created by the heavy-ion collisions as explained in the next part or realised in compact stellar objects~\cite{Fukushima:2020yzx}.

\begin{figure}[htbp!]
\centering
\includegraphics[width=0.7\linewidth]{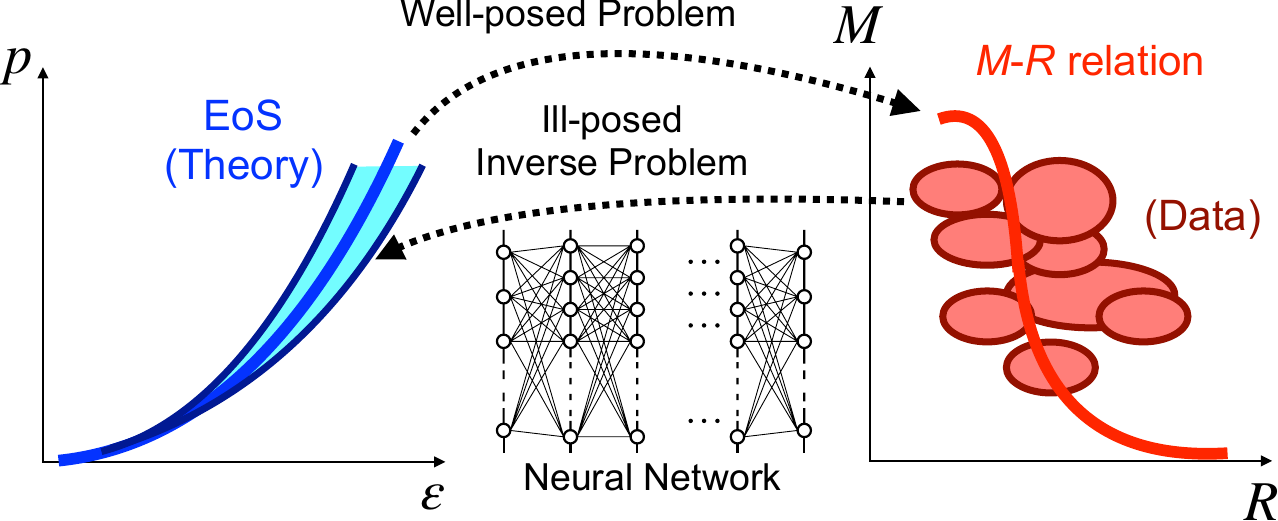}
\caption{Conceptual figure for the ill-posed inverse problem from observational data to the theory.  In the context of the neutron star physics, in the theoretical level, the mapping between the equation of state (EoS) and the mass-radius ($M$-$R$) relation is well-posed, but the inferred likely EoS has a probability distribution reflecting the quality and quantity of data.}
\label{fig:neutronstar}
\end{figure}

Neutron stars form a self-bound system assisted by the gravitational force.  The range of typical neutron star mass is $(1.4$-$2)$$M_\odot$ where $M_\odot$ is the solar mass.  One can imagine how dense the neutron star is from the comparison that the sun has its mass $M_\odot$ within the solar radius $\simeq 7\times 10^5\;\mathrm{km}$, while the neutron star radius is only $\simeq 10\;\mathrm{km}$.  It is a crucial question to identify internal structures of the neutron stars from theoretical predictions and experimental observations\cite{Baym:2017whm,Yunes:2022ldq}.  As long as General Relativity and QCD are absolutely correct theories, in principle, the bulk properties of the neutron stars are uniquely solved; this is a well-posed problem (see Fig.~\ref{fig:neutronstar}). Nevertheless, knowing the theories that lead to the correct answer is still different from knowing the answer itself.  Because of the sign problem in finite-density QCD, the lattice simulation does not work for the neutron star problem, and the essential information derived from QCD has huge uncertainties.  Specifically, a relation between the pressure and the density called the equation of state (EoS) is indispensable for the hydrostatic equations which describe the balance between the inward gravitational force and the outward pressure derivative.

For a given QCD EoS, one can deal with equations in General Relativity to thoroughly fix the distribution of the masses and the radii of the neutron stars, i.e., the $M$-$R$ relation that in principle draws a single curve.  Conversely, one can reconstruct the QCD EoS from the experimentally observed $M$-$R$ relation if it is well constrained.  This is a typical inverse problem!  However, the quality and quantity of observed data is insufficient.  Thus, we should cope with the ill-posed inverse problem of inferring the EoS from the limited observation.  To this end, what one could do the best is to estimate the most likely candidate among all EoS possibilities.  A naive strategy along these lines would be the following: random EoSs are generated with an assumed prior distribution, and the likelihood to get the observation from these EoSs is weighted, and then the posterior distribution is quantified~\cite{Steiner:2012xt}.  These procedures are systematised in a form of the Bayesian analysis~\cite{Ozel:2015fia}; see also a recent work including latest observation~\cite{Brandes:2022nxa}.  Alternatively, the supervised model of the neural network is good at this kind of inference~\cite{Fujimoto:2019hxv}.  The implementation of neural network with automatic differentiation has also be attempted~\cite{Soma:2022qnv,Soma:2022vbb}. A hybrid of the Bayesian and neural network is also proposed~\cite{Carvalho:2023ele,Carvalho:2024kgf}.  

The neutron star problem has an advantage in the context of physics-driven learning.  The observational data follows the underlying physics principles, and as stated above, the theories are trustworthy by themselves; in other words, the well-posed limit in the problem is guaranteed.  This is an ideal platform under theoretical control to test new ideas.  For example, the performance study of data augmentation has been discussed in this way~\cite{Fujimoto:2021zas}, the redesign of activation functions within neural networks represented EoSs has been employed to meet the causality principle~\cite{Han:2021kjx}.

\subsection*{Heavy-Ion-Collisions}

Relativistic heavy-ion collisions (HICs) provide the unique chance to create and explore the extreme state of QCD matter in the terrestrial laboratories, e.g., Relativistic Heavy-Ion Collider and Large Hadron Collider. In HICs, charged ions are accelerated to nearly the speed of light, collide, and create thousands of hadrons, photons, and leptons tracked by the detectors. Analyses of the final-state particles conjectured that a new state of matter -- Quark Gluon Plasma (QGP) is formed in the early stage of the collision. The formation of QGP is an important examination of the quark model and the QCD theory, as it is in the color deconfinement phase --- where quarks and gluons can travel distances that greatly exceed the size of hadrons. Soon after their formation in HICs, the QGP droplets then expand --- driven by the pressure gradient --- cool down, and hadronize. An inevitable challenge in HICs is inversely revealing the underlying QCD physics because of the highly complex and superfast evolving collision dynamics. Many physical uncertainties are involved in the state-of-the-art theoretical simulation for the collision dynamics, e.g., fluctuations of the initial state as well as QCD matter's bulk and transport properties, with entangled influence to different experimental observables~\cite{Bass:2017zyn}. DL techniques, boosted by the physics knowledges, have been shown to be useful in such inverse problems in HICs.

\begin{figure}[htbp!]
\centering
\includegraphics[width=0.9\linewidth]{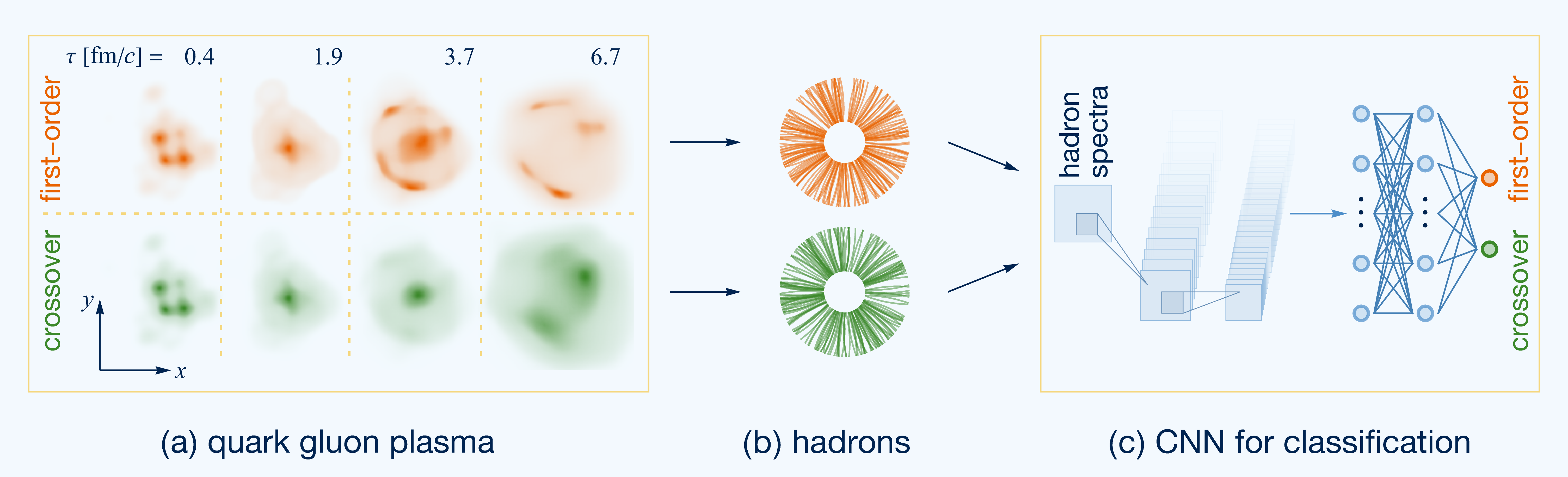}
\caption{A schematic flow chart for QCD transition binary classification with CNNs using final particle spectra from HICs as input. (a) Quark gluon plasma (QGP). The hydrodynamic evolution of QGP with the crossover and first-order phase transition encoded, respectively.
(b) Hadrons. The final observations collected by detectors are signals of different hadrons. (c) Convolutional neural networks (CNNs). CNNs are suitable for classification task in identifying the phase transition signals from hadron spectra which are preprocessed as image-type data.}
\label{fig:hic}
\end{figure}
A central task of HICs is to determine the QCD EoS, especially the relevant phase structure.
DL, specifically CNN, was used to construct a direct inverse mapping from final state spectra in HICs to QCD bulk properties recently\cite{Pang:2016vdc}, see the flow chart demonstrated in Figure.~\ref{fig:hic}. This exploratory study demonstrated that information about QCD transition--as one crucial feature embedded in the QCD EoS--can survive the relativistic hydrodynamic evolution into the event-by-event final state pion's spectra, and the decipherer from the CNN can decode back this information with robustness to other influencing uncertainties including initial fluctuations and shear viscosities. Such a way for inverse HICs was deepened further to confront the non-equilibrium phase transition scenarios and afterburner hadronic cascade\cite{Du:2020poe, Steinheimer:2021hoc, Jiang:2021gsw}, and also to include experiment detector effects where the hits and tracks of particles from the detector in the form of PointCloud --- defined as an unordered list of points with their attributes like momentum and charge --- is taken as direct input\cite{OmanaKuttan:2020btb}.

The extraction of QCD EoS is also benefited from physics-driven Bayesian analyses, in which knowledge are encoded in the priors and model emulator.
Inference of temperature dependent speed of sound from high-energy HICs\cite{Pratt:2015zsa} provided the first experimental evidence of a smooth transition between QGP and hadronic phase at zero baryon density. The Bayesian analysis was successfully applied to extract the density-dependent QCD EoS\cite{OmanaKuttan:2022aml} as well as bulk and shear viscosities\cite{Bernhard:2019bmu, JETSCAPE:2020shq, Nijs:2020ors}. It shall be interesting to note that good consistency was obtained between such data-driven inference of viscosities and those computed from a quasi-particle model with temperature-dependent mass extracted from a physics-driven learning to match the lattice QCD calculation of the EoS\cite{Li:2022ozl}.

Initial state of HICs is sensitive to lots of interesting physics, including nuclear structure and saturation of gluon distribution at small Bjorken-$x$. 
However, not all information of the initial state can be reconstructed based on final state experimental observables, since entropy is produced during the QGP evolution, which means the loss of information. The inversion to recover even though part of the initial state information can reveal correlations or collision patterns within the evolution dynamics. 
For instance, attempts invoking Bayesian inference have been made to probe the nuclear structure\cite{Cheng:2023ucp} or neutron skin\cite{Giacalone:2023cet}.
The impact parameter is a simple but essential initial state information. It governs the event geometry and volume estimation.
Different from early attempts using ML algorithms like MLP or support vector machine to determine the impact parameter from conventional observables\cite{David:1994qc, Bass:1996ez, DeSanctis:2009zzb}, a recent PointNet-based development\cite{qi2017pointnet,OmanaKuttan:2020brq} took advantages of the fact that the detector output in HICs inherently has a point cloud structure. Such an advance allows for a real-time end-to-end eventwise impact parameter determination for the next generation low-energy HICs to perform at the CBM detector of FAIR. 

\section*{Conclusions and outlook}

In conclusion, exploring QCD physics as inverse problems presents numerous challenges that drive innovative approaches to decode complex phenomena. Machine learning(ML) techniques, especially statistical inference and deep learning, have significantly advanced our ability to tackle these problems. Physics-driven learning ensures solutions align with physical realities, enhancing precision and relevance. Embedding symmetries, principles, and physical equations into deep model design opens new avenues for reliable physics-driven learning. The new paradigm has reformed first-principle calculations of Lattice QCD and significantly influenced investigations in hadron physics, neutron stars, and heavy-ion collisions. Although advancements in holographic QCD, jet physics, and Electron Ion Collider (EIC) physics are not detailed due to space constraints~\cite{Feickert:2021ajf}, it is noteworthy that anti-de Sitter/conformal field theory (AdS/CFT) correspondence has intrinsic connections with deep learning~\cite{Hashimoto:2018ftp}. This relationship offers new ways to develop holographic theories using neural networks within physics-driven frameworks~\cite{Hashimoto:2018bnb,Cai:2024eqa}.

Although it is still challenging to provide a general solution for solving inverse problems, the physics-driven learning is turning out to be particularly useful for utilising physics knowledge to regularise. This approach can significantly enhance the analysis of complex systems where traditional data-driven models fall short~\cite{kadambi2023incorporating}, such as in climate science for better predicting weather patterns~\cite{reichstein2019deep}, in biophysics for understanding active matter behaviours~\cite{cichos2020machine}, and in astrophysics for understanding cosmic phenomena~\cite{Huerta:2019rtg}. Furthermore, physics-driven learning can reduce the computational cost and data requirements by narrowing the parameter space, making it a powerful tool for advancing research in various domains of physical science.

Looking forward, the synergy between physics and ML promises continued advancements in both fields. The mathematical frameworks and problem-solving techniques of physics inspire ML algorithms and models, driving innovations that create more physically meaningful generative models and align deep learning models with the physical world. As ML evolves, its application to solving inverse problems will yield more precise and insightful results, thereby facilitating a deeper understanding of the fundamental interactions that underpin the universe.

\bibliography{ref}

\begin{thebibliography}{100}
\urlstyle{rm}
\expandafter\ifx\csname url\endcsname\relax
  \def\url#1{\texttt{#1}}\fi
\expandafter\ifx\csname urlprefix\endcsname\relax\def\urlprefix{URL }\fi
\expandafter\ifx\csname doiprefix\endcsname\relax\def\doiprefix{DOI: }\fi
\providecommand{\bibinfo}[2]{#2}
\providecommand{\eprint}[2][]{\url{#2}}

\bibitem{Gross:2022hyw}
\bibinfo{author}{Gross, F.} \emph{et~al.}
\newblock \bibinfo{journal}{\bibinfo{title}{{50 Years of Quantum Chromodynamics}}}.
\newblock {\emph{\JournalTitle{Eur. Phys. J. C}}} \textbf{\bibinfo{volume}{83}}, \bibinfo{pages}{1125}, \doiprefix\url{10.1140/epjc/s10052-023-11949-2} (\bibinfo{year}{2023}).
\newblock \eprint{2212.11107}.

\bibitem{Gattringer:2010zz}
\bibinfo{author}{Gattringer, C.} \& \bibinfo{author}{Lang, C.~B.}
\newblock \emph{\bibinfo{title}{{Quantum chromodynamics on the lattice}}}, vol. \bibinfo{volume}{788} (\bibinfo{publisher}{Springer}, \bibinfo{address}{Berlin}, \bibinfo{year}{2010}).

\bibitem{Baym:2017whm}
\bibinfo{author}{Baym, G.} \emph{et~al.}
\newblock \bibinfo{journal}{\bibinfo{title}{{From hadrons to quarks in neutron stars: a review}}}.
\newblock {\emph{\JournalTitle{Rept. Prog. Phys.}}} \textbf{\bibinfo{volume}{81}}, \bibinfo{pages}{056902}, \doiprefix\url{10.1088/1361-6633/aaae14} (\bibinfo{year}{2018}).
\newblock \eprint{1707.04966}.

\bibitem{Yagi:2005yb}
\bibinfo{author}{Yagi, K.}, \bibinfo{author}{Hatsuda, T.} \& \bibinfo{author}{Miake, Y.}
\newblock \emph{\bibinfo{title}{{Quark-gluon plasma: From big bang to little bang}}}, vol.~\bibinfo{volume}{23} (\bibinfo{publisher}{Cambridge University Press}, \bibinfo{year}{2005}).

\bibitem{tanaka2021deep}
\bibinfo{author}{Tanaka, A.}, \bibinfo{author}{Tomiya, A.} \& \bibinfo{author}{Hashimoto, K.}
\newblock \emph{\bibinfo{title}{Deep learning and physics}}, vol.~\bibinfo{volume}{1} (\bibinfo{publisher}{Springer}, \bibinfo{year}{2021}).

\bibitem{Zhou:2023pti}
\bibinfo{author}{Zhou, K.}, \bibinfo{author}{Wang, L.}, \bibinfo{author}{Pang, L.-G.} \& \bibinfo{author}{Shi, S.}
\newblock \bibinfo{journal}{\bibinfo{title}{{Exploring QCD matter in extreme conditions with Machine Learning}}}.
\newblock {\emph{\JournalTitle{Prog. Part. Nucl. Phys.}}} \textbf{\bibinfo{volume}{135}}, \bibinfo{pages}{104084}, \doiprefix\url{10.1016/j.ppnp.2023.104084} (\bibinfo{year}{2024}).
\newblock \eprint{2303.15136}.

\bibitem{kaipio2006statistical}
\bibinfo{author}{Kaipio, J.} \& \bibinfo{author}{Somersalo, E.}
\newblock \emph{\bibinfo{title}{Statistical and computational inverse problems}}, vol. \bibinfo{volume}{160} (\bibinfo{publisher}{Springer Science \& Business Media}, \bibinfo{year}{2006}).

\bibitem{Asakawa:2000tr}
\bibinfo{author}{Asakawa, M.}, \bibinfo{author}{Hatsuda, T.} \& \bibinfo{author}{Nakahara, Y.}
\newblock \bibinfo{journal}{\bibinfo{title}{{Maximum entropy analysis of the spectral functions in lattice QCD}}}.
\newblock {\emph{\JournalTitle{Prog. Part. Nucl. Phys.}}} \textbf{\bibinfo{volume}{46}}, \bibinfo{pages}{459--508}, \doiprefix\url{10.1016/S0146-6410(01)00150-8} (\bibinfo{year}{2001}).
\newblock \eprint{hep-lat/0011040}.

\bibitem{Yunes:2022ldq}
\bibinfo{author}{Yunes, N.}, \bibinfo{author}{Miller, M.~C.} \& \bibinfo{author}{Yagi, K.}
\newblock \bibinfo{journal}{\bibinfo{title}{{Gravitational-wave and X-ray probes of the neutron star equation of state}}}.
\newblock {\emph{\JournalTitle{Nature Rev. Phys.}}} \textbf{\bibinfo{volume}{4}}, \bibinfo{pages}{237--246}, \doiprefix\url{10.1038/s42254-022-00420-y} (\bibinfo{year}{2022}).
\newblock \eprint{2202.04117}.

\bibitem{Boehnlein:2021eym}
\bibinfo{author}{Boehnlein, A.} \emph{et~al.}
\newblock \bibinfo{journal}{\bibinfo{title}{{Colloquium: Machine learning in nuclear physics}}}.
\newblock {\emph{\JournalTitle{Rev. Mod. Phys.}}} \textbf{\bibinfo{volume}{94}}, \bibinfo{pages}{031003}, \doiprefix\url{10.1103/RevModPhys.94.031003} (\bibinfo{year}{2022}).
\newblock \eprint{2112.02309}.

\bibitem{LeCun:2015pmr}
\bibinfo{author}{LeCun, Y.}, \bibinfo{author}{Bengio, Y.} \& \bibinfo{author}{Hinton, G.}
\newblock \bibinfo{journal}{\bibinfo{title}{{Deep learning}}}.
\newblock {\emph{\JournalTitle{Nature}}} \textbf{\bibinfo{volume}{521}}, \bibinfo{pages}{436--444}, \doiprefix\url{10.1038/nature14539} (\bibinfo{year}{2015}).

\bibitem{bishop2023deep}
\bibinfo{author}{Bishop, C.~M.} \& \bibinfo{author}{Bishop, H.}
\newblock \emph{\bibinfo{title}{Deep learning: Foundations and concepts}} (\bibinfo{publisher}{Springer Nature}, \bibinfo{year}{2023}).

\bibitem{Raissi:2017zsi}
\bibinfo{author}{Raissi, M.}, \bibinfo{author}{Perdikaris, P.} \& \bibinfo{author}{Karniadakis, G.~E.}
\newblock \bibinfo{journal}{\bibinfo{title}{{Physics Informed Deep Learning (Part I): Data-driven Solutions of Nonlinear Partial Differential Equations}}}.
\newblock {\emph{\JournalTitle{J. Comput. Phys.}}} \textbf{\bibinfo{volume}{378}}, \bibinfo{pages}{686--707}, \doiprefix\url{10.1016/j.jcp.2018.10.045} (\bibinfo{year}{2019}).
\newblock \eprint{1711.10561}.

\bibitem{2021arXiv210905237T}
\bibinfo{author}{{Thuerey}, N.} \emph{et~al.}
\newblock \bibinfo{journal}{\bibinfo{title}{{Physics-based Deep Learning}}}.
\newblock {\emph{\JournalTitle{arXiv e-prints}}} \bibinfo{pages}{arXiv:2109.05237}, \doiprefix\url{10.48550/arXiv.2109.05237} (\bibinfo{year}{2021}).
\newblock \eprint{2109.05237}.

\bibitem{Carleo:2019ptp}
\bibinfo{author}{Carleo, G.} \emph{et~al.}
\newblock \bibinfo{journal}{\bibinfo{title}{{Machine learning and the physical sciences}}}.
\newblock {\emph{\JournalTitle{Rev. Mod. Phys.}}} \textbf{\bibinfo{volume}{91}}, \bibinfo{pages}{045002}, \doiprefix\url{10.1103/RevModPhys.91.045002} (\bibinfo{year}{2019}).
\newblock \eprint{1903.10563}.

\bibitem{lecun2006tutorial}
\bibinfo{author}{LeCun, Y.}, \bibinfo{author}{Chopra, S.}, \bibinfo{author}{Hadsell, R.}, \bibinfo{author}{Ranzato, M.} \& \bibinfo{author}{Huang, F.}
\newblock \bibinfo{journal}{\bibinfo{title}{A tutorial on energy-based learning}}.
\newblock {\emph{\JournalTitle{Predicting structured data}}} \textbf{\bibinfo{volume}{1}} (\bibinfo{year}{2006}).

\bibitem{Jaynes:1957zza}
\bibinfo{author}{Jaynes, E.~T.}
\newblock \bibinfo{journal}{\bibinfo{title}{{Information Theory and Statistical Mechanics}}}.
\newblock {\emph{\JournalTitle{Phys. Rev.}}} \textbf{\bibinfo{volume}{106}}, \bibinfo{pages}{620--630}, \doiprefix\url{10.1103/PhysRev.106.620} (\bibinfo{year}{1957}).

\bibitem{DBLP:journals/corr/Sohl-DicksteinW15}
\bibinfo{author}{Sohl{-}Dickstein, J.}, \bibinfo{author}{Weiss, E.~A.}, \bibinfo{author}{Maheswaranathan, N.} \& \bibinfo{author}{Ganguli, S.}
\newblock \bibinfo{journal}{\bibinfo{title}{Deep unsupervised learning using nonequilibrium thermodynamics}}.
\newblock {\emph{\JournalTitle{CoRR}}} \textbf{\bibinfo{volume}{abs/1503.03585}} (\bibinfo{year}{2015}).
\newblock \eprint{1503.03585}.

\bibitem{muller1995neural}
\bibinfo{author}{M{\"u}ller, B.}, \bibinfo{author}{Reinhardt, J.} \& \bibinfo{author}{Strickland, M.~T.}
\newblock \emph{\bibinfo{title}{Neural networks: an introduction}} (\bibinfo{publisher}{Springer Berlin, Heidelberg}, \bibinfo{year}{1995}).

\bibitem{murphy2012machine}
\bibinfo{author}{Murphy, K.~P.}
\newblock \emph{\bibinfo{title}{Machine learning: a probabilistic perspective}} (\bibinfo{publisher}{MIT press}, \bibinfo{year}{2012}).

\bibitem{7949028}
\bibinfo{author}{Jin, K.~H.}, \bibinfo{author}{McCann, M.~T.}, \bibinfo{author}{Froustey, E.} \& \bibinfo{author}{Unser, M.}
\newblock \bibinfo{journal}{\bibinfo{title}{Deep convolutional neural network for inverse problems in imaging}}.
\newblock {\emph{\JournalTitle{IEEE Transactions on Image Processing}}} \textbf{\bibinfo{volume}{26}}, \bibinfo{pages}{4509--4522}, \doiprefix\url{10.1109/TIP.2017.2713099} (\bibinfo{year}{2017}).

\bibitem{AMIN2007491}
\bibinfo{author}{Amin, G.~R.} \& \bibinfo{author}{Emrouznejad, A.}
\newblock \bibinfo{journal}{\bibinfo{title}{Inverse forecasting: A new approach for predictive modeling}}.
\newblock {\emph{\JournalTitle{Computers and Industrial Engineering}}} \textbf{\bibinfo{volume}{53}}, \bibinfo{pages}{491--498}, \doiprefix\url{https://doi.org/10.1016/j.cie.2007.05.007} (\bibinfo{year}{2007}).

\bibitem{behrmann2019invertible}
\bibinfo{author}{Behrmann, J.}, \bibinfo{author}{Grathwohl, W.}, \bibinfo{author}{Chen, R.~T.}, \bibinfo{author}{Duvenaud, D.} \& \bibinfo{author}{Jacobsen, J.-H.}
\newblock \bibinfo{title}{Invertible residual networks}.
\newblock In \emph{\bibinfo{booktitle}{International conference on machine learning}}, \bibinfo{pages}{573--582} (\bibinfo{organization}{PMLR}, \bibinfo{year}{2019}).

\bibitem{qi2017pointnet}
\bibinfo{author}{Qi, C.~R.}, \bibinfo{author}{Su, H.}, \bibinfo{author}{Mo, K.} \& \bibinfo{author}{Guibas, L.~J.}
\newblock \bibinfo{title}{Pointnet: Deep learning on point sets for 3d classification and segmentation}.
\newblock In \emph{\bibinfo{booktitle}{Proceedings of the IEEE conference on computer vision and pattern recognition}}, \bibinfo{pages}{652--660} (\bibinfo{year}{2017}).

\bibitem{Tomiya:2023jdy}
\bibinfo{author}{Tomiya, A.} \& \bibinfo{author}{Nagai, Y.}
\newblock \bibinfo{journal}{\bibinfo{title}{{Equivariant transformer is all you need}}}.
\newblock {\emph{\JournalTitle{PoS}}} \textbf{\bibinfo{volume}{LATTICE2023}}, \bibinfo{pages}{001}, \doiprefix\url{10.22323/1.453.0001} (\bibinfo{year}{2024}).
\newblock \eprint{2310.13222}.

\bibitem{raissi2019physics}
\bibinfo{author}{Raissi, M.}, \bibinfo{author}{Perdikaris, P.} \& \bibinfo{author}{Karniadakis, G.~E.}
\newblock \bibinfo{journal}{\bibinfo{title}{Physics-informed neural networks: A deep learning framework for solving forward and inverse problems involving nonlinear partial differential equations}}.
\newblock {\emph{\JournalTitle{Journal of Computational physics}}} \textbf{\bibinfo{volume}{378}}, \bibinfo{pages}{686--707} (\bibinfo{year}{2019}).

\bibitem{2019arXiv190408991M}
\bibinfo{author}{{Mattheakis}, M.}, \bibinfo{author}{{Protopapas}, P.}, \bibinfo{author}{{Sondak}, D.}, \bibinfo{author}{{Di Giovanni}, M.} \& \bibinfo{author}{{Kaxiras}, E.}
\newblock \bibinfo{journal}{\bibinfo{title}{{Physical Symmetries Embedded in Neural Networks}}}.
\newblock {\emph{\JournalTitle{arXiv e-prints}}} \bibinfo{pages}{arXiv:1904.08991}, \doiprefix\url{10.48550/arXiv.1904.08991} (\bibinfo{year}{2019}).
\newblock \eprint{1904.08991}.

\bibitem{zhang1988shift}
\bibinfo{author}{Zhang, W.}, \bibinfo{author}{Tanida, J.}, \bibinfo{author}{Itoh, K.} \& \bibinfo{author}{Ichioka, Y.}
\newblock \bibinfo{title}{Shift-invariant pattern recognition neural network and its optical architecture}.
\newblock In \emph{\bibinfo{booktitle}{Proceedings of annual conference of the Japan Society of Applied Physics}}, vol. \bibinfo{volume}{564} (\bibinfo{organization}{Montreal, CA}, \bibinfo{year}{1988}).

\bibitem{2022arXiv220709453G}
\bibinfo{author}{{Geiger}, M.} \& \bibinfo{author}{{Smidt}, T.}
\newblock \bibinfo{journal}{\bibinfo{title}{{e3nn: Euclidean Neural Networks}}}.
\newblock {\emph{\JournalTitle{arXiv e-prints}}} \bibinfo{pages}{arXiv:2207.09453}, \doiprefix\url{10.48550/arXiv.2207.09453} (\bibinfo{year}{2022}).
\newblock \eprint{2207.09453}.

\bibitem{Shlomi_2021}
\bibinfo{author}{Shlomi, J.}, \bibinfo{author}{Battaglia, P.} \& \bibinfo{author}{Vlimant, J.-R.}
\newblock \bibinfo{journal}{\bibinfo{title}{Graph neural networks in particle physics}}.
\newblock {\emph{\JournalTitle{Machine Learning: Science and Technology}}} \textbf{\bibinfo{volume}{2}}, \bibinfo{pages}{021001}, \doiprefix\url{10.1088/2632-2153/abbf9a} (\bibinfo{year}{2020}).

\bibitem{Kanwar:2020xzo}
\bibinfo{author}{Kanwar, G.} \emph{et~al.}
\newblock \bibinfo{journal}{\bibinfo{title}{{Equivariant flow-based sampling for lattice gauge theory}}}.
\newblock {\emph{\JournalTitle{Phys. Rev. Lett.}}} \textbf{\bibinfo{volume}{125}}, \bibinfo{pages}{121601}, \doiprefix\url{10.1103/PhysRevLett.125.121601} (\bibinfo{year}{2020}).
\newblock \eprint{2003.06413}.

\bibitem{Favoni:2020reg}
\bibinfo{author}{Favoni, M.}, \bibinfo{author}{Ipp, A.}, \bibinfo{author}{M\"uller, D.~I.} \& \bibinfo{author}{Schuh, D.}
\newblock \bibinfo{journal}{\bibinfo{title}{{Lattice Gauge Equivariant Convolutional Neural Networks}}}.
\newblock {\emph{\JournalTitle{Phys. Rev. Lett.}}} \textbf{\bibinfo{volume}{128}}, \bibinfo{pages}{032003}, \doiprefix\url{10.1103/PhysRevLett.128.032003} (\bibinfo{year}{2022}).
\newblock \eprint{2012.12901}.

\bibitem{Cranmer:2023xbe}
\bibinfo{author}{Cranmer, K.}, \bibinfo{author}{Kanwar, G.}, \bibinfo{author}{Racani\`ere, S.}, \bibinfo{author}{Rezende, D.~J.} \& \bibinfo{author}{Shanahan, P.~E.}
\newblock \bibinfo{journal}{\bibinfo{title}{{Advances in machine-learning-based sampling motivated by lattice quantum chromodynamics}}}.
\newblock {\emph{\JournalTitle{Nature Rev. Phys.}}} \textbf{\bibinfo{volume}{5}}, \bibinfo{pages}{526--535}, \doiprefix\url{10.1038/s42254-023-00616-w} (\bibinfo{year}{2023}).
\newblock \eprint{2309.01156}.

\bibitem{Han:2021kjx}
\bibinfo{author}{Han, M.-Z.}, \bibinfo{author}{Jiang, J.-L.}, \bibinfo{author}{Tang, S.-P.} \& \bibinfo{author}{Fan, Y.-Z.}
\newblock \bibinfo{journal}{\bibinfo{title}{{Bayesian Nonparametric Inference of the Neutron Star Equation of State via a Neural Network}}}.
\newblock {\emph{\JournalTitle{Astrophys. J.}}} \textbf{\bibinfo{volume}{919}}, \bibinfo{pages}{11}, \doiprefix\url{10.3847/1538-4357/ac11f8} (\bibinfo{year}{2021}).
\newblock \eprint{2103.05408}.

\bibitem{Shi:2022yqw}
\bibinfo{author}{Shi, S.}, \bibinfo{author}{Wang, L.} \& \bibinfo{author}{Zhou, K.}
\newblock \bibinfo{journal}{\bibinfo{title}{{Rethinking the ill-posedness of the spectral function reconstruction \textemdash{} Why is it fundamentally hard and how Artificial Neural Networks can help}}}.
\newblock {\emph{\JournalTitle{Comput. Phys. Commun.}}} \textbf{\bibinfo{volume}{282}}, \bibinfo{pages}{108547}, \doiprefix\url{10.1016/j.cpc.2022.108547} (\bibinfo{year}{2023}).
\newblock \eprint{2201.02564}.

\bibitem{karniadakis2021physics}
\bibinfo{author}{Karniadakis, G.~E.} \emph{et~al.}
\newblock \bibinfo{journal}{\bibinfo{title}{Physics-informed machine learning}}.
\newblock {\emph{\JournalTitle{Nature Reviews Physics}}} \textbf{\bibinfo{volume}{3}}, \bibinfo{pages}{422--440} (\bibinfo{year}{2021}).

\bibitem{baydin2018automatic}
\bibinfo{author}{Baydin, A.~G.}, \bibinfo{author}{Pearlmutter, B.~A.}, \bibinfo{author}{Radul, A.~A.} \& \bibinfo{author}{Siskind, J.~M.}
\newblock \bibinfo{journal}{\bibinfo{title}{Automatic differentiation in machine learning: a survey}}.
\newblock {\emph{\JournalTitle{Journal of machine learning research}}} \textbf{\bibinfo{volume}{18}}, \bibinfo{pages}{1--43} (\bibinfo{year}{2018}).

\bibitem{Cranmer:2019eaq}
\bibinfo{author}{Cranmer, K.}, \bibinfo{author}{Brehmer, J.} \& \bibinfo{author}{Louppe, G.}
\newblock \bibinfo{journal}{\bibinfo{title}{{The frontier of simulation-based inference}}}.
\newblock {\emph{\JournalTitle{Proc. Nat. Acad. Sci.}}} \textbf{\bibinfo{volume}{117}}, \bibinfo{pages}{30055--30062}, \doiprefix\url{10.1073/pnas.1912789117} (\bibinfo{year}{2020}).
\newblock \eprint{1911.01429}.

\bibitem{wang2023scientific}
\bibinfo{author}{Wang, H.} \emph{et~al.}
\newblock \bibinfo{journal}{\bibinfo{title}{Scientific discovery in the age of artificial intelligence}}.
\newblock {\emph{\JournalTitle{Nature}}} \textbf{\bibinfo{volume}{620}}, \bibinfo{pages}{47--60} (\bibinfo{year}{2023}).

\bibitem{jalali2022physics}
\bibinfo{author}{Jalali, B.}, \bibinfo{author}{Zhou, Y.}, \bibinfo{author}{Kadambi, A.} \& \bibinfo{author}{Roychowdhury, V.}
\newblock \bibinfo{journal}{\bibinfo{title}{Physics-ai symbiosis}}.
\newblock {\emph{\JournalTitle{Machine Learning: Science and Technology}}} \textbf{\bibinfo{volume}{3}}, \bibinfo{pages}{041001} (\bibinfo{year}{2022}).

\bibitem{Aarts:2015tyj}
\bibinfo{author}{Aarts, G.}
\newblock \bibinfo{journal}{\bibinfo{title}{{Introductory lectures on lattice QCD at nonzero baryon number}}}.
\newblock {\emph{\JournalTitle{J. Phys. Conf. Ser.}}} \textbf{\bibinfo{volume}{706}}, \bibinfo{pages}{022004}, \doiprefix\url{10.1088/1742-6596/706/2/022004} (\bibinfo{year}{2016}).
\newblock \eprint{1512.05145}.

\bibitem{Boyda:2022nmh}
\bibinfo{author}{Boyda, D.} \emph{et~al.}
\newblock \bibinfo{title}{{Applications of Machine Learning to Lattice Quantum Field Theory}}.
\newblock In \emph{\bibinfo{booktitle}{{Snowmass 2021}}} (\bibinfo{year}{2022}).
\newblock \eprint{2202.05838}.

\bibitem{Hasenfratz:1993sp}
\bibinfo{author}{Hasenfratz, P.} \& \bibinfo{author}{Niedermayer, F.}
\newblock \bibinfo{journal}{\bibinfo{title}{{Perfect lattice action for asymptotically free theories}}}.
\newblock {\emph{\JournalTitle{Nucl. Phys. B}}} \textbf{\bibinfo{volume}{414}}, \bibinfo{pages}{785--814}, \doiprefix\url{10.1016/0550-3213(94)90261-5} (\bibinfo{year}{1994}).
\newblock \eprint{hep-lat/9308004}.

\bibitem{DeGrand:1995ji}
\bibinfo{author}{DeGrand, T.~A.}, \bibinfo{author}{Hasenfratz, A.}, \bibinfo{author}{Hasenfratz, P.} \& \bibinfo{author}{Niedermayer, F.}
\newblock \bibinfo{journal}{\bibinfo{title}{{The Classically perfect fixed point action for SU(3) gauge theory}}}.
\newblock {\emph{\JournalTitle{Nucl. Phys. B}}} \textbf{\bibinfo{volume}{454}}, \bibinfo{pages}{587--614}, \doiprefix\url{10.1016/0550-3213(95)00458-5} (\bibinfo{year}{1995}).
\newblock \eprint{hep-lat/9506030}.

\bibitem{Shanahan:2018vcv}
\bibinfo{author}{Shanahan, P.~E.}, \bibinfo{author}{Trewartha, A.} \& \bibinfo{author}{Detmold, W.}
\newblock \bibinfo{journal}{\bibinfo{title}{{Machine learning action parameters in lattice quantum chromodynamics}}}.
\newblock {\emph{\JournalTitle{Phys. Rev. D}}} \textbf{\bibinfo{volume}{97}}, \bibinfo{pages}{094506}, \doiprefix\url{10.1103/PhysRevD.97.094506} (\bibinfo{year}{2018}).
\newblock \eprint{1801.05784}.

\bibitem{Nagai:2020jar}
\bibinfo{author}{Nagai, Y.}, \bibinfo{author}{Tanaka, A.} \& \bibinfo{author}{Tomiya, A.}
\newblock \bibinfo{journal}{\bibinfo{title}{{Self-learning Monte Carlo for non-Abelian gauge theory with dynamical fermions}}}.
\newblock {\emph{\JournalTitle{Phys. Rev. D}}} \textbf{\bibinfo{volume}{107}}, \bibinfo{pages}{054501}, \doiprefix\url{10.1103/PhysRevD.107.054501} (\bibinfo{year}{2023}).
\newblock \eprint{2010.11900}.

\bibitem{Blucher:2020mjt}
\bibinfo{author}{Bl\"ucher, S.}, \bibinfo{author}{Kades, L.}, \bibinfo{author}{Pawlowski, J.~M.}, \bibinfo{author}{Strodthoff, N.} \& \bibinfo{author}{Urban, J.~M.}
\newblock \bibinfo{journal}{\bibinfo{title}{{Towards novel insights in lattice field theory with explainable machine learning}}}.
\newblock {\emph{\JournalTitle{Phys. Rev. D}}} \textbf{\bibinfo{volume}{101}}, \bibinfo{pages}{094507}, \doiprefix\url{10.1103/PhysRevD.101.094507} (\bibinfo{year}{2020}).
\newblock \eprint{2003.01504}.

\bibitem{Holland:2023ews}
\bibinfo{author}{Holland, K.}, \bibinfo{author}{Ipp, A.}, \bibinfo{author}{M\"uller, D.~I.} \& \bibinfo{author}{Wenger, U.}
\newblock \bibinfo{journal}{\bibinfo{title}{{Fixed point actions from convolutional neural networks}}}.
\newblock {\emph{\JournalTitle{PoS}}} \textbf{\bibinfo{volume}{LATTICE2023}}, \bibinfo{pages}{038}, \doiprefix\url{10.22323/1.453.0038} (\bibinfo{year}{2024}).
\newblock \eprint{2311.17816}.

\bibitem{Holland:2024muu}
\bibinfo{author}{Holland, K.}, \bibinfo{author}{Ipp, A.}, \bibinfo{author}{M\"uller, D.~I.} \& \bibinfo{author}{Wenger, U.}
\newblock \bibinfo{journal}{\bibinfo{title}{{Machine learning a fixed point action for SU(3) gauge theory with a gauge equivariant convolutional neural network}}}.
\newblock {\emph{\JournalTitle{arXiv preprint}}}  (\bibinfo{year}{2024}).
\newblock \eprint{2401.06481}.

\bibitem{Zhou:2018ill}
\bibinfo{author}{Zhou, K.}, \bibinfo{author}{Endr\H{o}di, G.}, \bibinfo{author}{Pang, L.-G.} \& \bibinfo{author}{St\"ocker, H.}
\newblock \bibinfo{journal}{\bibinfo{title}{{Regressive and generative neural networks for scalar field theory}}}.
\newblock {\emph{\JournalTitle{Phys. Rev. D}}} \textbf{\bibinfo{volume}{100}}, \bibinfo{pages}{011501}, \doiprefix\url{10.1103/PhysRevD.100.011501} (\bibinfo{year}{2019}).
\newblock \eprint{1810.12879}.

\bibitem{Pawlowski:2018qxs}
\bibinfo{author}{Pawlowski, J.~M.} \& \bibinfo{author}{Urban, J.~M.}
\newblock \bibinfo{journal}{\bibinfo{title}{{Reducing Autocorrelation Times in Lattice Simulations with Generative Adversarial Networks}}}.
\newblock {\emph{\JournalTitle{Mach. Learn. Sci. Tech.}}} \textbf{\bibinfo{volume}{1}}, \bibinfo{pages}{045011}, \doiprefix\url{10.1088/2632-2153/abae73} (\bibinfo{year}{2020}).
\newblock \eprint{1811.03533}.

\bibitem{Kanwar:2024ujc}
\bibinfo{author}{Kanwar, G.}
\newblock \bibinfo{title}{{Flow-based sampling for lattice field theories}}.
\newblock In \emph{\bibinfo{booktitle}{{40th International Symposium on Lattice Field Theory}}} (\bibinfo{year}{2024}).
\newblock \eprint{2401.01297}.

\bibitem{DBLP:journals/corr/abs-2011-13456}
\bibinfo{author}{Song, Y.} \emph{et~al.}
\newblock \bibinfo{journal}{\bibinfo{title}{Score-based generative modeling through stochastic differential equations}}.
\newblock {\emph{\JournalTitle{CoRR}}} \textbf{\bibinfo{volume}{abs/2011.13456}} (\bibinfo{year}{2020}).
\newblock \eprint{2011.13456}.

\bibitem{Parisi:1980ys}
\bibinfo{author}{Parisi, G.} \& \bibinfo{author}{Wu, Y.~S.}
\newblock \bibinfo{journal}{\bibinfo{title}{{Perturbation theory without gauge fixing}}}.
\newblock {\emph{\JournalTitle{Sci. China, A}}} \textbf{\bibinfo{volume}{24}}, \bibinfo{pages}{483} (\bibinfo{year}{1980}).

\bibitem{Wang:2023exq}
\bibinfo{author}{Wang, L.}, \bibinfo{author}{Aarts, G.} \& \bibinfo{author}{Zhou, K.}
\newblock \bibinfo{journal}{\bibinfo{title}{{Diffusion models as stochastic quantization in lattice field theory}}}.
\newblock {\emph{\JournalTitle{JHEP}}} \textbf{\bibinfo{volume}{05}}, \bibinfo{pages}{060}, \doiprefix\url{10.1007/JHEP05(2024)060} (\bibinfo{year}{2024}).
\newblock \eprint{2309.17082}.

\bibitem{Wang:2023sry}
\bibinfo{author}{Wang, L.}, \bibinfo{author}{Aarts, G.} \& \bibinfo{author}{Zhou, K.}
\newblock \bibinfo{title}{{Generative Diffusion Models for Lattice Field Theory}}.
\newblock In \emph{\bibinfo{booktitle}{{37th Conference on Neural Information Processing Systems}}} (\bibinfo{year}{2023}).
\newblock \eprint{2311.03578}.

\bibitem{Zhu:2024kiu}
\bibinfo{author}{Zhu, Q.}, \bibinfo{author}{Aarts, G.}, \bibinfo{author}{Wang, W.}, \bibinfo{author}{Zhou, K.} \& \bibinfo{author}{Wang, L.}
\newblock \bibinfo{title}{{Diffusion models for lattice gauge field simulations}}.
\newblock In \emph{\bibinfo{booktitle}{{38th conference on Neural Information Processing Systems}}} (\bibinfo{year}{2024}).
\newblock \eprint{2410.19602}.

\bibitem{Hirono:2024zyg}
\bibinfo{author}{Hirono, Y.}, \bibinfo{author}{Tanaka, A.} \& \bibinfo{author}{Fukushima, K.}
\newblock \bibinfo{journal}{\bibinfo{title}{{Understanding Diffusion Models by Feynman's Path Integral}}}.
\newblock {\emph{\JournalTitle{arXiv preprint}}}  (\bibinfo{year}{2024}).
\newblock \eprint{2403.11262}.

\bibitem{Cotler:2023lem}
\bibinfo{author}{Cotler, J.} \& \bibinfo{author}{Rezchikov, S.}
\newblock \bibinfo{title}{{Renormalizing Diffusion Models}} (\bibinfo{year}{2023}).
\newblock \eprint{2308.12355}.

\bibitem{PhysRevLett.89.275701}
\bibinfo{author}{Ron, D.}, \bibinfo{author}{Swendsen, R.~H.} \& \bibinfo{author}{Brandt, A.}
\newblock \bibinfo{journal}{\bibinfo{title}{Inverse monte carlo renormalization group transformations for critical phenomena}}.
\newblock {\emph{\JournalTitle{Phys. Rev. Lett.}}} \textbf{\bibinfo{volume}{89}}, \bibinfo{pages}{275701}, \doiprefix\url{10.1103/PhysRevLett.89.275701} (\bibinfo{year}{2002}).

\bibitem{Bachtis:2021eww}
\bibinfo{author}{Bachtis, D.}, \bibinfo{author}{Aarts, G.}, \bibinfo{author}{Di~Renzo, F.} \& \bibinfo{author}{Lucini, B.}
\newblock \bibinfo{journal}{\bibinfo{title}{{Inverse Renormalization Group in Quantum Field Theory}}}.
\newblock {\emph{\JournalTitle{Phys. Rev. Lett.}}} \textbf{\bibinfo{volume}{128}}, \bibinfo{pages}{081603}, \doiprefix\url{10.1103/PhysRevLett.128.081603} (\bibinfo{year}{2022}).
\newblock \eprint{2107.00466}.

\bibitem{Bachtis:2023ykk}
\bibinfo{author}{Bachtis, D.}
\newblock \bibinfo{journal}{\bibinfo{title}{{Inverse Renormalization Group of Disordered Systems}}}.
\newblock {\emph{\JournalTitle{arXiv preprint}}}  (\bibinfo{year}{2023}).
\newblock \eprint{2310.12631}.

\bibitem{Lehner:2023bba}
\bibinfo{author}{Lehner, C.} \& \bibinfo{author}{Wettig, T.}
\newblock \bibinfo{journal}{\bibinfo{title}{{Gauge-equivariant neural networks as preconditioners in lattice QCD}}}.
\newblock {\emph{\JournalTitle{Phys. Rev. D}}} \textbf{\bibinfo{volume}{108}}, \bibinfo{pages}{034503}, \doiprefix\url{10.1103/PhysRevD.108.034503} (\bibinfo{year}{2023}).
\newblock \eprint{2302.05419}.

\bibitem{Aronsson:2023rli}
\bibinfo{author}{Aronsson, J.}, \bibinfo{author}{M\"uller, D.~I.} \& \bibinfo{author}{Schuh, D.}
\newblock \bibinfo{journal}{\bibinfo{title}{{Geometrical aspects of lattice gauge equivariant convolutional neural networks}}}.
\newblock {\emph{\JournalTitle{arXiv preprint}}}  (\bibinfo{year}{2023}).
\newblock \eprint{2303.11448}.

\bibitem{Lehner:2023prf}
\bibinfo{author}{Lehner, C.} \& \bibinfo{author}{Wettig, T.}
\newblock \bibinfo{journal}{\bibinfo{title}{{Gauge-equivariant pooling layers for preconditioners in lattice QCD}}}.
\newblock {\emph{\JournalTitle{arXiv preprint}}}  (\bibinfo{year}{2023}).
\newblock \eprint{2304.10438}.

\bibitem{DBLP:journals/corr/CohenW16}
\bibinfo{author}{Cohen, T.~S.} \& \bibinfo{author}{Welling, M.}
\newblock \bibinfo{journal}{\bibinfo{title}{Group equivariant convolutional networks}}.
\newblock {\emph{\JournalTitle{CoRR}}} \textbf{\bibinfo{volume}{abs/1602.07576}} (\bibinfo{year}{2016}).
\newblock \eprint{1602.07576}.

\bibitem{DBLP:journals/corr/abs-1902-04615}
\bibinfo{author}{Cohen, T.~S.}, \bibinfo{author}{Weiler, M.}, \bibinfo{author}{Kicanaoglu, B.} \& \bibinfo{author}{Welling, M.}
\newblock \bibinfo{journal}{\bibinfo{title}{Gauge equivariant convolutional networks and the icosahedral {CNN}}}.
\newblock {\emph{\JournalTitle{CoRR}}} \textbf{\bibinfo{volume}{abs/1902.04615}} (\bibinfo{year}{2019}).
\newblock \eprint{1902.04615}.

\bibitem{Carrasquilla_2017}
\bibinfo{author}{Carrasquilla, J.} \& \bibinfo{author}{Melko, R.~G.}
\newblock \bibinfo{journal}{\bibinfo{title}{Machine learning phases of matter}}.
\newblock {\emph{\JournalTitle{Nature Physics}}} \textbf{\bibinfo{volume}{13}}, \bibinfo{pages}{431–434}, \doiprefix\url{10.1038/nphys4035} (\bibinfo{year}{2017}).

\bibitem{Wetzel:2017ooo}
\bibinfo{author}{Wetzel, S.~J.} \& \bibinfo{author}{Scherzer, M.}
\newblock \bibinfo{journal}{\bibinfo{title}{{Machine Learning of Explicit Order Parameters: From the Ising Model to SU(2) Lattice Gauge Theory}}}.
\newblock {\emph{\JournalTitle{Phys. Rev. B}}} \textbf{\bibinfo{volume}{96}}, \bibinfo{pages}{184410}, \doiprefix\url{10.1103/PhysRevB.96.184410} (\bibinfo{year}{2017}).
\newblock \eprint{1705.05582}.

\bibitem{Boyda:2020nfh}
\bibinfo{author}{Boyda, D.~L.} \emph{et~al.}
\newblock \bibinfo{journal}{\bibinfo{title}{{Finding the deconfinement temperature in lattice Yang-Mills theories from outside the scaling window with machine learning}}}.
\newblock {\emph{\JournalTitle{Phys. Rev. D}}} \textbf{\bibinfo{volume}{103}}, \bibinfo{pages}{014509}, \doiprefix\url{10.1103/PhysRevD.103.014509} (\bibinfo{year}{2021}).
\newblock \eprint{2009.10971}.

\bibitem{lee2018deep}
\bibinfo{author}{Lee, J.} \emph{et~al.}
\newblock \bibinfo{title}{Deep neural networks as gaussian processes} (\bibinfo{year}{2018}).
\newblock \eprint{1711.00165}.

\bibitem{Halverson:2020trp}
\bibinfo{author}{Halverson, J.}, \bibinfo{author}{Maiti, A.} \& \bibinfo{author}{Stoner, K.}
\newblock \bibinfo{journal}{\bibinfo{title}{{Neural Networks and Quantum Field Theory}}}.
\newblock {\emph{\JournalTitle{Mach. Learn. Sci. Tech.}}} \textbf{\bibinfo{volume}{2}}, \bibinfo{pages}{035002}, \doiprefix\url{10.1088/2632-2153/abeca3} (\bibinfo{year}{2021}).
\newblock \eprint{2008.08601}.

\bibitem{Bachtis:2021xoh}
\bibinfo{author}{Bachtis, D.}, \bibinfo{author}{Aarts, G.} \& \bibinfo{author}{Lucini, B.}
\newblock \bibinfo{journal}{\bibinfo{title}{{Quantum field-theoretic machine learning}}}.
\newblock {\emph{\JournalTitle{Phys. Rev. D}}} \textbf{\bibinfo{volume}{103}}, \bibinfo{pages}{074510}, \doiprefix\url{10.1103/PhysRevD.103.074510} (\bibinfo{year}{2021}).
\newblock \eprint{2102.09449}.

\bibitem{Aarts:2023uwt}
\bibinfo{author}{Aarts, G.}, \bibinfo{author}{Lucini, B.} \& \bibinfo{author}{Park, C.}
\newblock \bibinfo{journal}{\bibinfo{title}{{Scalar field restricted Boltzmann machine as an ultraviolet regulator}}}.
\newblock {\emph{\JournalTitle{Phys. Rev. D}}} \textbf{\bibinfo{volume}{109}}, \bibinfo{pages}{034521}, \doiprefix\url{10.1103/PhysRevD.109.034521} (\bibinfo{year}{2024}).
\newblock \eprint{2309.15002}.

\bibitem{Aarts:2024wxi}
\bibinfo{author}{Aarts, G.}, \bibinfo{author}{Lucini, B.} \& \bibinfo{author}{Park, C.}
\newblock \bibinfo{title}{{Stochastic weight matrix dynamics during learning and Dyson Brownian motion}} (\bibinfo{year}{2024}).
\newblock \eprint{2407.16427}.

\bibitem{Rothkopf:2019ipj}
\bibinfo{author}{Rothkopf, A.}
\newblock \bibinfo{journal}{\bibinfo{title}{{Heavy Quarkonium in Extreme Conditions}}}.
\newblock {\emph{\JournalTitle{Phys. Rept.}}} \textbf{\bibinfo{volume}{858}}, \bibinfo{pages}{1--117}, \doiprefix\url{10.1016/j.physrep.2020.02.006} (\bibinfo{year}{2020}).
\newblock \eprint{1912.02253}.

\bibitem{Wang:2021jou}
\bibinfo{author}{Wang, L.}, \bibinfo{author}{Shi, S.} \& \bibinfo{author}{Zhou, K.}
\newblock \bibinfo{journal}{\bibinfo{title}{{Reconstructing spectral functions via automatic differentiation}}}.
\newblock {\emph{\JournalTitle{Phys. Rev. D}}} \textbf{\bibinfo{volume}{106}}, \bibinfo{pages}{L051502}, \doiprefix\url{10.1103/PhysRevD.106.L051502} (\bibinfo{year}{2022}).
\newblock \eprint{2111.14760}.

\bibitem{Guo:2017jvc}
\bibinfo{author}{Guo, F.-K.} \emph{et~al.}
\newblock \bibinfo{journal}{\bibinfo{title}{{Hadronic molecules}}}.
\newblock {\emph{\JournalTitle{Rev. Mod. Phys.}}} \textbf{\bibinfo{volume}{90}}, \bibinfo{pages}{015004}, \doiprefix\url{10.1103/RevModPhys.90.015004} (\bibinfo{year}{2018}).
\newblock \bibinfo{note}{[Erratum: Rev.Mod.Phys. 94, 029901 (2022)]}, \eprint{1705.00141}.

\bibitem{Sombillo:2020ccg}
\bibinfo{author}{Sombillo, D. L.~B.}, \bibinfo{author}{Ikeda, Y.}, \bibinfo{author}{Sato, T.} \& \bibinfo{author}{Hosaka, A.}
\newblock \bibinfo{journal}{\bibinfo{title}{{Classifying the pole of an amplitude using a deep neural network}}}.
\newblock {\emph{\JournalTitle{Phys. Rev. D}}} \textbf{\bibinfo{volume}{102}}, \bibinfo{pages}{016024}, \doiprefix\url{10.1103/PhysRevD.102.016024} (\bibinfo{year}{2020}).
\newblock \eprint{2003.10770}.

\bibitem{Sombillo:2021rxv}
\bibinfo{author}{Sombillo, D. L.~B.}, \bibinfo{author}{Ikeda, Y.}, \bibinfo{author}{Sato, T.} \& \bibinfo{author}{Hosaka, A.}
\newblock \bibinfo{journal}{\bibinfo{title}{{Model independent analysis of coupled-channel scattering: A deep learning approach}}}.
\newblock {\emph{\JournalTitle{Phys. Rev. D}}} \textbf{\bibinfo{volume}{104}}, \bibinfo{pages}{036001}, \doiprefix\url{10.1103/PhysRevD.104.036001} (\bibinfo{year}{2021}).
\newblock \eprint{2105.04898}.

\bibitem{JPAC:2021rxu}
\bibinfo{author}{Albaladejo, M.} \emph{et~al.}
\newblock \bibinfo{journal}{\bibinfo{title}{{Novel approaches in hadron spectroscopy}}}.
\newblock {\emph{\JournalTitle{Prog. Part. Nucl. Phys.}}} \textbf{\bibinfo{volume}{127}}, \bibinfo{pages}{103981}, \doiprefix\url{10.1016/j.ppnp.2022.103981} (\bibinfo{year}{2022}).
\newblock \eprint{2112.13436}.

\bibitem{Ng:2021ibr}
\bibinfo{author}{Ng, L.} \emph{et~al.}
\newblock \bibinfo{journal}{\bibinfo{title}{{Deep learning exotic hadrons}}}.
\newblock {\emph{\JournalTitle{Phys. Rev. D}}} \textbf{\bibinfo{volume}{105}}, \bibinfo{pages}{L091501}, \doiprefix\url{10.1103/PhysRevD.105.L091501} (\bibinfo{year}{2022}).
\newblock \eprint{2110.13742}.

\bibitem{Keeble:2019bkv}
\bibinfo{author}{Keeble, J. W.~T.} \& \bibinfo{author}{Rios, A.}
\newblock \bibinfo{journal}{\bibinfo{title}{{Machine learning the deuteron}}}.
\newblock {\emph{\JournalTitle{Phys. Lett. B}}} \textbf{\bibinfo{volume}{809}}, \bibinfo{pages}{135743}, \doiprefix\url{10.1016/j.physletb.2020.135743} (\bibinfo{year}{2020}).
\newblock \eprint{1911.13092}.

\bibitem{Adams:2020aax}
\bibinfo{author}{Adams, C.}, \bibinfo{author}{Carleo, G.}, \bibinfo{author}{Lovato, A.} \& \bibinfo{author}{Rocco, N.}
\newblock \bibinfo{journal}{\bibinfo{title}{{Variational Monte Carlo Calculations of A\ensuremath{\leq}4 Nuclei with an Artificial Neural-Network Correlator Ansatz}}}.
\newblock {\emph{\JournalTitle{Phys. Rev. Lett.}}} \textbf{\bibinfo{volume}{127}}, \bibinfo{pages}{022502}, \doiprefix\url{10.1103/PhysRevLett.127.022502} (\bibinfo{year}{2021}).
\newblock \eprint{2007.14282}.

\bibitem{Ishii:2006ec}
\bibinfo{author}{Ishii, N.}, \bibinfo{author}{Aoki, S.} \& \bibinfo{author}{Hatsuda, T.}
\newblock \bibinfo{journal}{\bibinfo{title}{{The Nuclear Force from Lattice QCD}}}.
\newblock {\emph{\JournalTitle{Phys. Rev. Lett.}}} \textbf{\bibinfo{volume}{99}}, \bibinfo{pages}{022001}, \doiprefix\url{10.1103/PhysRevLett.99.022001} (\bibinfo{year}{2007}).
\newblock \eprint{nucl-th/0611096}.

\bibitem{Aoki:2012tk}
\bibinfo{author}{Aoki, S.} \emph{et~al.}
\newblock \bibinfo{journal}{\bibinfo{title}{{Lattice QCD approach to Nuclear Physics}}}.
\newblock {\emph{\JournalTitle{PTEP}}} \textbf{\bibinfo{volume}{2012}}, \bibinfo{pages}{01A105}, \doiprefix\url{10.1093/ptep/pts010} (\bibinfo{year}{2012}).
\newblock \eprint{1206.5088}.

\bibitem{Aoki:2023qih}
\bibinfo{author}{Aoki, S.} \& \bibinfo{author}{Doi, T.}
\newblock \emph{\bibinfo{title}{{Lattice QCD and Baryon-Baryon Interactions}}}, \bibinfo{pages}{1--31} (\bibinfo{publisher}{{Springer}}, \bibinfo{year}{2023}).
\newblock \eprint{2402.11759}.

\bibitem{ALICE:2020mfd}
\bibinfo{author}{Collaboration, A.} \emph{et~al.}
\newblock \bibinfo{journal}{\bibinfo{title}{{Unveiling the strong interaction among hadrons at the LHC}}}.
\newblock {\emph{\JournalTitle{Nature}}} \textbf{\bibinfo{volume}{588}}, \bibinfo{pages}{232--238}, \doiprefix\url{10.1038/s41586-020-3001-6} (\bibinfo{year}{2020}).
\newblock \bibinfo{note}{[Erratum: Nature 590, E13 (2021)]}, \eprint{2005.11495}.

\bibitem{Lyu:2024kwz}
\bibinfo{author}{Lyu, Y.} \emph{et~al.}
\newblock \bibinfo{journal}{\bibinfo{title}{{Doubly charmed tetraquark $T_{cc}^+$ in (2+1)-flavor QCD near physical point}}}.
\newblock {\emph{\JournalTitle{PoS}}} \textbf{\bibinfo{volume}{LATTICE2023}}, \bibinfo{pages}{077}, \doiprefix\url{10.22323/1.453.0077} (\bibinfo{year}{2024}).
\newblock \eprint{2401.13917}.

\bibitem{Shi:2021qri}
\bibinfo{author}{Shi, S.}, \bibinfo{author}{Zhou, K.}, \bibinfo{author}{Zhao, J.}, \bibinfo{author}{Mukherjee, S.} \& \bibinfo{author}{Zhuang, P.}
\newblock \bibinfo{journal}{\bibinfo{title}{{Heavy quark potential in the quark-gluon plasma: Deep neural network meets lattice quantum chromodynamics}}}.
\newblock {\emph{\JournalTitle{Phys. Rev. D}}} \textbf{\bibinfo{volume}{105}}, \bibinfo{pages}{014017}, \doiprefix\url{10.1103/PhysRevD.105.014017} (\bibinfo{year}{2022}).
\newblock \eprint{2105.07862}.

\bibitem{Wang:2024ykk}
\bibinfo{author}{Wang, L.}, \bibinfo{author}{Doi, T.}, \bibinfo{author}{Hatsuda, T.} \& \bibinfo{author}{Lyu, Y.}
\newblock \bibinfo{title}{{Building Hadron Potentials from Lattice QCD with Deep Neural Networks}} (\bibinfo{year}{2024}).
\newblock \eprint{2410.03082}.

\bibitem{Fukushima:2020yzx}
\bibinfo{author}{Fukushima, K.}, \bibinfo{author}{Mohanty, B.} \& \bibinfo{author}{Xu, N.}
\newblock \bibinfo{journal}{\bibinfo{title}{{Little-Bang and Femto-Nova in Nucleus-Nucleus Collisions}}}.
\newblock {\emph{\JournalTitle{AAPPS Bull.}}} \textbf{\bibinfo{volume}{31}}, \bibinfo{pages}{1}, \doiprefix\url{10.1007/s43673-021-00002-7} (\bibinfo{year}{2021}).
\newblock \eprint{2009.03006}.

\bibitem{Steiner:2012xt}
\bibinfo{author}{Steiner, A.~W.}, \bibinfo{author}{Lattimer, J.~M.} \& \bibinfo{author}{Brown, E.~F.}
\newblock \bibinfo{journal}{\bibinfo{title}{{The Neutron Star Mass-Radius Relation and the Equation of State of Dense Matter}}}.
\newblock {\emph{\JournalTitle{Astrophys. J. Lett.}}} \textbf{\bibinfo{volume}{765}}, \bibinfo{pages}{L5}, \doiprefix\url{10.1088/2041-8205/765/1/L5} (\bibinfo{year}{2013}).
\newblock \eprint{1205.6871}.

\bibitem{Ozel:2015fia}
\bibinfo{author}{Ozel, F.} \emph{et~al.}
\newblock \bibinfo{journal}{\bibinfo{title}{{The Dense Matter Equation of State from Neutron Star Radius and Mass Measurements}}}.
\newblock {\emph{\JournalTitle{Astrophys. J.}}} \textbf{\bibinfo{volume}{820}}, \bibinfo{pages}{28}, \doiprefix\url{10.3847/0004-637X/820/1/28} (\bibinfo{year}{2016}).
\newblock \eprint{1505.05155}.

\bibitem{Brandes:2022nxa}
\bibinfo{author}{Brandes, L.}, \bibinfo{author}{Weise, W.} \& \bibinfo{author}{Kaiser, N.}
\newblock \bibinfo{journal}{\bibinfo{title}{{Inference of the sound speed and related properties of neutron stars}}}.
\newblock {\emph{\JournalTitle{Phys. Rev. D}}} \textbf{\bibinfo{volume}{107}}, \bibinfo{pages}{014011}, \doiprefix\url{10.1103/PhysRevD.107.014011} (\bibinfo{year}{2023}).
\newblock \eprint{2208.03026}.

\bibitem{Fujimoto:2019hxv}
\bibinfo{author}{Fujimoto, Y.}, \bibinfo{author}{Fukushima, K.} \& \bibinfo{author}{Murase, K.}
\newblock \bibinfo{journal}{\bibinfo{title}{{Mapping neutron star data to the equation of state using the deep neural network}}}.
\newblock {\emph{\JournalTitle{Phys. Rev. D}}} \textbf{\bibinfo{volume}{101}}, \bibinfo{pages}{054016}, \doiprefix\url{10.1103/PhysRevD.101.054016} (\bibinfo{year}{2020}).
\newblock \eprint{1903.03400}.

\bibitem{Soma:2022qnv}
\bibinfo{author}{Soma, S.}, \bibinfo{author}{Wang, L.}, \bibinfo{author}{Shi, S.}, \bibinfo{author}{St\"ocker, H.} \& \bibinfo{author}{Zhou, K.}
\newblock \bibinfo{journal}{\bibinfo{title}{{Neural network reconstruction of the dense matter equation of state from neutron star observables}}}.
\newblock {\emph{\JournalTitle{JCAP}}} \textbf{\bibinfo{volume}{08}}, \bibinfo{pages}{071}, \doiprefix\url{10.1088/1475-7516/2022/08/071} (\bibinfo{year}{2022}).
\newblock \eprint{2201.01756}.

\bibitem{Soma:2022vbb}
\bibinfo{author}{Soma, S.}, \bibinfo{author}{Wang, L.}, \bibinfo{author}{Shi, S.}, \bibinfo{author}{St\"ocker, H.} \& \bibinfo{author}{Zhou, K.}
\newblock \bibinfo{journal}{\bibinfo{title}{{Reconstructing the neutron star equation of state from observational data via automatic differentiation}}}.
\newblock {\emph{\JournalTitle{Phys. Rev. D}}} \textbf{\bibinfo{volume}{107}}, \bibinfo{pages}{083028}, \doiprefix\url{10.1103/PhysRevD.107.083028} (\bibinfo{year}{2023}).
\newblock \eprint{2209.08883}.

\bibitem{Carvalho:2023ele}
\bibinfo{author}{Carvalho, V.}, \bibinfo{author}{Ferreira, M.}, \bibinfo{author}{Malik, T.} \& \bibinfo{author}{Provid\^encia, C.}
\newblock \bibinfo{journal}{\bibinfo{title}{{Decoding neutron star observations: Revealing composition through Bayesian neural networks}}}.
\newblock {\emph{\JournalTitle{Phys. Rev. D}}} \textbf{\bibinfo{volume}{108}}, \bibinfo{pages}{043031}, \doiprefix\url{10.1103/PhysRevD.108.043031} (\bibinfo{year}{2023}).
\newblock \eprint{2306.06929}.

\bibitem{Carvalho:2024kgf}
\bibinfo{author}{Carvalho, V.}, \bibinfo{author}{Ferreira, M.} \& \bibinfo{author}{Provid\^encia, C.}
\newblock \bibinfo{journal}{\bibinfo{title}{{From NS observations to nuclear matter properties: a machine learning approach}}}.
\newblock {\emph{\JournalTitle{arXiv preprint}}}  (\bibinfo{year}{2024}).
\newblock \eprint{2401.05770}.

\bibitem{Fujimoto:2021zas}
\bibinfo{author}{Fujimoto, Y.}, \bibinfo{author}{Fukushima, K.} \& \bibinfo{author}{Murase, K.}
\newblock \bibinfo{journal}{\bibinfo{title}{{Extensive Studies of the Neutron Star Equation of State from the Deep Learning Inference with the Observational Data Augmentation}}}.
\newblock {\emph{\JournalTitle{JHEP}}} \textbf{\bibinfo{volume}{03}}, \bibinfo{pages}{273}, \doiprefix\url{10.1007/JHEP03(2021)273} (\bibinfo{year}{2021}).
\newblock \eprint{2101.08156}.

\bibitem{Bass:2017zyn}
\bibinfo{author}{Bass, S.~A.}, \bibinfo{author}{Bernhard, J.~E.} \& \bibinfo{author}{Moreland, J.~S.}
\newblock \bibinfo{journal}{\bibinfo{title}{{Determination of Quark-Gluon-Plasma Parameters from a Global Bayesian Analysis}}}.
\newblock {\emph{\JournalTitle{Nucl. Phys. A}}} \textbf{\bibinfo{volume}{967}}, \bibinfo{pages}{67--73}, \doiprefix\url{10.1016/j.nuclphysa.2017.05.052} (\bibinfo{year}{2017}).
\newblock \eprint{1704.07671}.

\bibitem{Pang:2016vdc}
\bibinfo{author}{Pang, L.-G.} \emph{et~al.}
\newblock \bibinfo{journal}{\bibinfo{title}{{An equation-of-state-meter of quantum chromodynamics transition from deep learning}}}.
\newblock {\emph{\JournalTitle{Nature Commun.}}} \textbf{\bibinfo{volume}{9}}, \bibinfo{pages}{210}, \doiprefix\url{10.1038/s41467-017-02726-3} (\bibinfo{year}{2018}).
\newblock \eprint{1612.04262}.

\bibitem{Du:2020poe}
\bibinfo{author}{Du, Y.-L.} \emph{et~al.}
\newblock \bibinfo{journal}{\bibinfo{title}{{Identifying the nature of the QCD transition in heavy-ion collisions with deep learning}}}.
\newblock {\emph{\JournalTitle{Nucl. Phys. A}}} \textbf{\bibinfo{volume}{1005}}, \bibinfo{pages}{121891}, \doiprefix\url{10.1016/j.nuclphysa.2020.121891} (\bibinfo{year}{2021}).
\newblock \eprint{2009.03059}.

\bibitem{Steinheimer:2021hoc}
\bibinfo{author}{Steinheimer, J.} \emph{et~al.}
\newblock \bibinfo{journal}{\bibinfo{title}{{A machine learning study on spinodal clumping in heavy ion collisions}}}.
\newblock {\emph{\JournalTitle{Nucl. Phys. A}}} \textbf{\bibinfo{volume}{1005}}, \bibinfo{pages}{121867}, \doiprefix\url{10.1016/j.nuclphysa.2020.121867} (\bibinfo{year}{2021}).

\bibitem{Jiang:2021gsw}
\bibinfo{author}{Jiang, L.}, \bibinfo{author}{Wang, L.} \& \bibinfo{author}{Zhou, K.}
\newblock \bibinfo{journal}{\bibinfo{title}{{Deep learning stochastic processes with QCD phase transition}}}.
\newblock {\emph{\JournalTitle{Phys. Rev. D}}} \textbf{\bibinfo{volume}{103}}, \bibinfo{pages}{116023}, \doiprefix\url{10.1103/PhysRevD.103.116023} (\bibinfo{year}{2021}).
\newblock \eprint{2103.04090}.

\bibitem{OmanaKuttan:2020btb}
\bibinfo{author}{Omana~Kuttan, M.}, \bibinfo{author}{Zhou, K.}, \bibinfo{author}{Steinheimer, J.}, \bibinfo{author}{Redelbach, A.} \& \bibinfo{author}{Stoecker, H.}
\newblock \bibinfo{journal}{\bibinfo{title}{{An equation-of-state-meter for CBM using PointNet}}}.
\newblock {\emph{\JournalTitle{JHEP}}} \textbf{\bibinfo{volume}{21}}, \bibinfo{pages}{184}, \doiprefix\url{10.1007/JHEP10(2021)184} (\bibinfo{year}{2020}).
\newblock \eprint{2107.05590}.

\bibitem{Pratt:2015zsa}
\bibinfo{author}{Pratt, S.}, \bibinfo{author}{Sangaline, E.}, \bibinfo{author}{Sorensen, P.} \& \bibinfo{author}{Wang, H.}
\newblock \bibinfo{journal}{\bibinfo{title}{{Constraining the Eq. of State of Super-Hadronic Matter from Heavy-Ion Collisions}}}.
\newblock {\emph{\JournalTitle{Phys. Rev. Lett.}}} \textbf{\bibinfo{volume}{114}}, \bibinfo{pages}{202301}, \doiprefix\url{10.1103/PhysRevLett.114.202301} (\bibinfo{year}{2015}).
\newblock \eprint{1501.04042}.

\bibitem{OmanaKuttan:2022aml}
\bibinfo{author}{Omana~Kuttan, M.}, \bibinfo{author}{Steinheimer, J.}, \bibinfo{author}{Zhou, K.} \& \bibinfo{author}{Stoecker, H.}
\newblock \bibinfo{journal}{\bibinfo{title}{{QCD Equation of State of Dense Nuclear Matter from a Bayesian Analysis of Heavy-Ion Collision Data}}}.
\newblock {\emph{\JournalTitle{Phys. Rev. Lett.}}} \textbf{\bibinfo{volume}{131}}, \bibinfo{pages}{202303}, \doiprefix\url{10.1103/PhysRevLett.131.202303} (\bibinfo{year}{2023}).
\newblock \eprint{2211.11670}.

\bibitem{Bernhard:2019bmu}
\bibinfo{author}{Bernhard, J.~E.}, \bibinfo{author}{Moreland, J.~S.} \& \bibinfo{author}{Bass, S.~A.}
\newblock \bibinfo{journal}{\bibinfo{title}{{Bayesian estimation of the specific shear and bulk viscosity of quark\textendash{}gluon plasma}}}.
\newblock {\emph{\JournalTitle{Nature Phys.}}} \textbf{\bibinfo{volume}{15}}, \bibinfo{pages}{1113--1117}, \doiprefix\url{10.1038/s41567-019-0611-8} (\bibinfo{year}{2019}).

\bibitem{JETSCAPE:2020shq}
\bibinfo{author}{Everett, D.} \emph{et~al.}
\newblock \bibinfo{journal}{\bibinfo{title}{{Phenomenological constraints on the transport properties of QCD matter with data-driven model averaging}}}.
\newblock {\emph{\JournalTitle{Phys. Rev. Lett.}}} \textbf{\bibinfo{volume}{126}}, \bibinfo{pages}{242301}, \doiprefix\url{10.1103/PhysRevLett.126.242301} (\bibinfo{year}{2021}).
\newblock \eprint{2010.03928}.

\bibitem{Nijs:2020ors}
\bibinfo{author}{Nijs, G.}, \bibinfo{author}{van~der Schee, W.}, \bibinfo{author}{G\"ursoy, U.} \& \bibinfo{author}{Snellings, R.}
\newblock \bibinfo{journal}{\bibinfo{title}{{Transverse Momentum Differential Global Analysis of Heavy-Ion Collisions}}}.
\newblock {\emph{\JournalTitle{Phys. Rev. Lett.}}} \textbf{\bibinfo{volume}{126}}, \bibinfo{pages}{202301}, \doiprefix\url{10.1103/PhysRevLett.126.202301} (\bibinfo{year}{2021}).
\newblock \eprint{2010.15130}.

\bibitem{Li:2022ozl}
\bibinfo{author}{Li, F.-P.}, \bibinfo{author}{L\"u, H.-L.}, \bibinfo{author}{Pang, L.-G.} \& \bibinfo{author}{Qin, G.-Y.}
\newblock \bibinfo{journal}{\bibinfo{title}{{Deep-learning quasi-particle masses from QCD equation of state}}}.
\newblock {\emph{\JournalTitle{Phys. Lett. B}}} \textbf{\bibinfo{volume}{844}}, \bibinfo{pages}{138088}, \doiprefix\url{10.1016/j.physletb.2023.138088} (\bibinfo{year}{2023}).
\newblock \eprint{2211.07994}.

\bibitem{Cheng:2023ucp}
\bibinfo{author}{Cheng, Y.-L.}, \bibinfo{author}{Shi, S.}, \bibinfo{author}{Ma, Y.-G.}, \bibinfo{author}{St\"ocker, H.} \& \bibinfo{author}{Zhou, K.}
\newblock \bibinfo{journal}{\bibinfo{title}{{Examination of nucleon distribution with Bayesian imaging for isobar collisions}}}.
\newblock {\emph{\JournalTitle{Phys. Rev. C}}} \textbf{\bibinfo{volume}{107}}, \bibinfo{pages}{064909}, \doiprefix\url{10.1103/PhysRevC.107.064909} (\bibinfo{year}{2023}).
\newblock \eprint{2301.03910}.

\bibitem{Giacalone:2023cet}
\bibinfo{author}{Giacalone, G.}, \bibinfo{author}{Nijs, G.} \& \bibinfo{author}{van~der Schee, W.}
\newblock \bibinfo{journal}{\bibinfo{title}{{Determination of the Neutron Skin of Pb208 from Ultrarelativistic Nuclear Collisions}}}.
\newblock {\emph{\JournalTitle{Phys. Rev. Lett.}}} \textbf{\bibinfo{volume}{131}}, \bibinfo{pages}{202302}, \doiprefix\url{10.1103/PhysRevLett.131.202302} (\bibinfo{year}{2023}).
\newblock \eprint{2305.00015}.

\bibitem{David:1994qc}
\bibinfo{author}{David, C.}, \bibinfo{author}{Freslier, M.} \& \bibinfo{author}{Aichelin, J.}
\newblock \bibinfo{journal}{\bibinfo{title}{{Impact parameter determination for heavy-ion collisions by use of a neural network}}}.
\newblock {\emph{\JournalTitle{Phys. Rev. C}}} \textbf{\bibinfo{volume}{51}}, \bibinfo{pages}{1453--1459}, \doiprefix\url{10.1103/PhysRevC.51.1453} (\bibinfo{year}{1995}).

\bibitem{Bass:1996ez}
\bibinfo{author}{Bass, S.~A.}, \bibinfo{author}{Bischoff, A.}, \bibinfo{author}{Maruhn, J.~A.}, \bibinfo{author}{Stoecker, H.} \& \bibinfo{author}{Greiner, W.}
\newblock \bibinfo{journal}{\bibinfo{title}{{Neural networks for impact parameter determination}}}.
\newblock {\emph{\JournalTitle{Phys. Rev. C}}} \textbf{\bibinfo{volume}{53}}, \bibinfo{pages}{2358--2363}, \doiprefix\url{10.1103/PhysRevC.53.2358} (\bibinfo{year}{1996}).
\newblock \eprint{nucl-th/9601024}.

\bibitem{DeSanctis:2009zzb}
\bibinfo{author}{De~Sanctis, J.} \emph{et~al.}
\newblock \bibinfo{journal}{\bibinfo{title}{{Classification of the impact parameter in nucleus-nucleus collisions by a support vector machine method}}}.
\newblock {\emph{\JournalTitle{J. Phys. G}}} \textbf{\bibinfo{volume}{36}}, \bibinfo{pages}{015101}, \doiprefix\url{10.1088/0954-3899/36/1/015101} (\bibinfo{year}{2009}).

\bibitem{OmanaKuttan:2020brq}
\bibinfo{author}{Omana~Kuttan, M.}, \bibinfo{author}{Steinheimer, J.}, \bibinfo{author}{Zhou, K.}, \bibinfo{author}{Redelbach, A.} \& \bibinfo{author}{Stoecker, H.}
\newblock \bibinfo{journal}{\bibinfo{title}{{A fast centrality-meter for heavy-ion collisions at the CBM experiment}}}.
\newblock {\emph{\JournalTitle{Phys. Lett. B}}} \textbf{\bibinfo{volume}{811}}, \bibinfo{pages}{135872}, \doiprefix\url{10.1016/j.physletb.2020.135872} (\bibinfo{year}{2020}).
\newblock \eprint{2009.01584}.

\bibitem{Feickert:2021ajf}
\bibinfo{author}{Feickert, M.} \& \bibinfo{author}{Nachman, B.}
\newblock \bibinfo{journal}{\bibinfo{title}{{A Living Review of Machine Learning for Particle Physics}}}.
\newblock {\emph{\JournalTitle{arXiv preprint}}}  (\bibinfo{year}{2021}).
\newblock \eprint{2102.02770}.

\bibitem{Hashimoto:2018ftp}
\bibinfo{author}{Hashimoto, K.}, \bibinfo{author}{Sugishita, S.}, \bibinfo{author}{Tanaka, A.} \& \bibinfo{author}{Tomiya, A.}
\newblock \bibinfo{journal}{\bibinfo{title}{{Deep learning and the AdS/CFT correspondence}}}.
\newblock {\emph{\JournalTitle{Phys. Rev. D}}} \textbf{\bibinfo{volume}{98}}, \bibinfo{pages}{046019}, \doiprefix\url{10.1103/PhysRevD.98.046019} (\bibinfo{year}{2018}).
\newblock \eprint{1802.08313}.

\bibitem{Hashimoto:2018bnb}
\bibinfo{author}{Hashimoto, K.}, \bibinfo{author}{Sugishita, S.}, \bibinfo{author}{Tanaka, A.} \& \bibinfo{author}{Tomiya, A.}
\newblock \bibinfo{journal}{\bibinfo{title}{{Deep Learning and Holographic QCD}}}.
\newblock {\emph{\JournalTitle{Phys. Rev. D}}} \textbf{\bibinfo{volume}{98}}, \bibinfo{pages}{106014}, \doiprefix\url{10.1103/PhysRevD.98.106014} (\bibinfo{year}{2018}).
\newblock \eprint{1809.10536}.

\bibitem{Cai:2024eqa}
\bibinfo{author}{Cai, R.-G.}, \bibinfo{author}{He, S.}, \bibinfo{author}{Li, L.} \& \bibinfo{author}{Zeng, H.-A.}
\newblock \bibinfo{journal}{\bibinfo{title}{{QCD Phase Diagram at finite Magnetic Field and Chemical Potential: A Holographic Approach Using Machine Learning}}}.
\newblock {\emph{\JournalTitle{arXiv preprint}}}  (\bibinfo{year}{2024}).
\newblock \eprint{2406.12772}.

\bibitem{kadambi2023incorporating}
\bibinfo{author}{Kadambi, A.}, \bibinfo{author}{de~Melo, C.}, \bibinfo{author}{Hsieh, C.-J.}, \bibinfo{author}{Srivastava, M.} \& \bibinfo{author}{Soatto, S.}
\newblock \bibinfo{journal}{\bibinfo{title}{Incorporating physics into data-driven computer vision}}.
\newblock {\emph{\JournalTitle{Nature Machine Intelligence}}} \textbf{\bibinfo{volume}{5}}, \bibinfo{pages}{572--580} (\bibinfo{year}{2023}).

\bibitem{reichstein2019deep}
\bibinfo{author}{Reichstein, M.} \emph{et~al.}
\newblock \bibinfo{journal}{\bibinfo{title}{Deep learning and process understanding for data-driven earth system science}}.
\newblock {\emph{\JournalTitle{Nature}}} \textbf{\bibinfo{volume}{566}}, \bibinfo{pages}{195--204} (\bibinfo{year}{2019}).

\bibitem{cichos2020machine}
\bibinfo{author}{Cichos, F.}, \bibinfo{author}{Gustavsson, K.}, \bibinfo{author}{Mehlig, B.} \& \bibinfo{author}{Volpe, G.}
\newblock \bibinfo{journal}{\bibinfo{title}{Machine learning for active matter}}.
\newblock {\emph{\JournalTitle{Nature Machine Intelligence}}} \textbf{\bibinfo{volume}{2}}, \bibinfo{pages}{94--103} (\bibinfo{year}{2020}).

\bibitem{Huerta:2019rtg}
\bibinfo{author}{Huerta, E.~A.} \emph{et~al.}
\newblock \bibinfo{journal}{\bibinfo{title}{{Enabling real-time multi-messenger astrophysics discoveries with deep learning}}}.
\newblock {\emph{\JournalTitle{Nature Rev. Phys.}}} \textbf{\bibinfo{volume}{1}}, \bibinfo{pages}{600--608}, \doiprefix\url{10.1038/s42254-019-0097-4} (\bibinfo{year}{2019}).
\newblock \eprint{1911.11779}.

\end{thebibliography}

\section*{Acknowledgements}
We thank
Len~Brandes, 
Yuki~Fujimoto, 
Syo~Kamata, 
David~I.~M\"uller, 
Koichi~Murase, 
Akinori~Tanaka 
for helpful discussions. 
We thank the DEEP-IN working group at RIKEN-iTHEMS, ECT* and the ExtreMe Matter Institute (EMMI) for their support in the preparation of this paper.
GA is supported by STFC Consolidated Grant ST/T000813/1.
KF is supported by Japan Society for the Promotion of Science (JSPS) KAKENHI Grant Nos.\ 22H01216 and 22H05118.
TH is supported by the Japan Science and Technology Agency (JST) as part of Adopting Sustainable Partnerships for Innovative Research Ecosystem (ASPIRE),
Grant No. JPMJAP2318.
SS acknowledges Tsinghua University under grant No. 53330500923.
LW thanks the National Natural Science Foundation of China(No.12147101) for supporting his visit to Fudan University.
KZ is supported by the CUHK-Shenzhen university development fund under grant No. UDF01003041 and UDF03003041, and Shenzhen Peacock fund under No. 2023TC0179.

\section*{Author contributions}
The authors contributed equally to all aspects of the article.

\section*{Competing interests}
The authors declare no competing interests.

\section*{Publisher’s note}
Springer Nature remains neutral with regard to jurisdictional claims in published maps and institutional affiliations.

\section*{Glossary terms}

\textbf{Artificial neural networks(ANNs)}: ANN is a model inspired by the structure and function of biological neural networks in human brains.

\textbf{Deep neural networks (DNNs):} Also named as Deep Net, a complex ANN with multiple layers, including input, output, and at least one hidden layer.

\textbf{Convolutional neural networks (CNNs):} CNNs excel with image, speech, and audio inputs. It consists of three main types of layers, convolutional layer, pooling layer, fully-connected layer.

\textbf{Recurrent neural networks (RNNs):} RNN is a bi-directional ANN, unlike the uni-directional feedforward network. It allows outputs from nodes to influence subsequent inputs to the same nodes.

\end{document}